\documentclass[
    reprint, twocolumn,
    aps,prb,10pt,amsmath,amssymb,
    longbibliography,superscriptaddress,
    showpacs,preprintnumbers,]{revtex4-2}

\usepackage[colorlinks=true,citecolor=blue,linkcolor=blue,urlcolor=blue]{hyperref}
\usepackage{oubraces}
\usepackage{bm,bbm}
\usepackage{mathtools}
\usepackage{amsmath, amsthm, amssymb, amsfonts}
\usepackage{bbold}
\usepackage{esvect}
\usepackage[dvipsnames]{xcolor}

\usepackage{epsfig}
\usepackage{array}
\usepackage{multirow}
\usepackage{graphicx,tabularx}
\usepackage[normalem]{ulem}

\usepackage{float}
\usepackage[section]{placeins}
\usepackage{afterpage}
\usepackage{xspace}

\usepackage{lipsum}

\usepackage{xifthen}

\makeatletter
\newcommand\footnoteref[1]{\protected@xdef\@thefnmark{\ref{#1}}\@footnotemark}
\makeatother

\newcommand{\nnnl}{\nonumber \\}

\newcommand{\qe}[0]{\textsc{Quantum ESPRESSO}\xspace}

\newcommand{\Eq}[1]{Eq.~\eqref{#1}}
\newcommand{\Eqs}[1]{Eqs.~\eqref{#1}}
\newcommand{\Equ}[1]{Equation~\eqref{#1}}
\newcommand{\Equs}[1]{Equations~\eqref{#1}}
\newcommand{\Sec}[1]{Sec.~\ref{#1}}

\newcommand{\Fig}[1]{Fig.~\ref{#1}}
\newcommand{\Figs}[1]{Figs.~\ref{#1}}
\newcommand{\Figu}[1]{Figure~\ref{#1}}
\newcommand{\Figus}[1]{Figures~\ref{#1}}

\newcommand{\mb}[1]{{\boldsymbol{\mathbf{#1}}}}

\newcommand{\bk}[0]{{\mb{k}}}

\newcommand{\bR}[0]{{\mb{R}}}
\newcommand{\bmR}[0]{{\mb{-R}}}
\newcommand{\bRp}[0]{{\mb{R'}}}
\newcommand{\br}[0]{{\mb{r}}}
\newcommand{\bb}[0]{{\mb{b}}}
\newcommand{\bd}[0]{{\mb{d}}}
\newcommand{\bkb}[0]{{\mb{k} + \mb{b}}}

\newcommand{\kb}[0]{{(\bk, \bb)}}
\newcommand{\nk}[0]{{n\bk}}
\newcommand{\mk}[0]{{m\bk}}

\newcommand{\ik}[0]{{i\bk}}
\newcommand{\jk}[0]{{j\bk}}

\newcommand{\jkb}[0]{{j\bkb}}

\newcommand{\nkb}[0]{{n\bkb}}
\newcommand{\iR}[0]{{i\bR}}
\newcommand{\jR}[0]{{j\bR}}

\newcommand{\jRp}[0]{{j\bRp}}
\newcommand{\iiR}[0]{{ii\bR}}
\newcommand{\ijR}[0]{{ij\bR}}

\newcommand{\veps}[0]{\varepsilon}

\newcommand{\dd}[0]{{\mathrm{d}}}

\renewcommand{\Im}{\operatorname{Im}}

\newcommand{\expikR}[0]{{e^{i\mb{k}\cdot\mb{R}}}}

\newcommand{\opmbp}[0]{\hat{\mb{p}}}

\newcommand{\ai}[0]{{\alpha_1}}
\newcommand{\aii}[0]{{\alpha_2}}
\newcommand{\an}[0]{{\alpha_n}}

\newcommand{\nint}{\!\int\!} 
\newcommand{\nbint}[2]{\!\int_{#1}^{#2}\!\!} 

\DeclarePairedDelimiterX\mel[3]{\langle}{\rangle}{#1 \delimsize\vert\mathopen{} #2 \delimsize\vert\mathopen{} #3}
\DeclarePairedDelimiterX\abs[1]{\lvert}{\rvert}{#1}
\newcommand{\expval}[1]{{\langle #1 \rangle}}
\newcommand{\ket}[1]{{\lvert #1 \rangle}}

\newcommand{\braket}[2]{{\langle #1 \vert #2 \rangle}}
\usepackage[utf8]{inputenc}
\usepackage{kotex}
\usepackage{orcidlink}
\usepackage{makecell}
\usepackage{comment}
\usepackage{cancel}
\usepackage{bbding}

\def\multiset#1#2{\ensuremath{\left(\kern-.3em\left(\genfrac{}{}{0pt}{}{#1}{#2}\right)\kern-.3em\right)}}

\ifx\ShowRevision\undefined
    
    \newcommand{\mydelete}[2]{}
    \newcommand{\mydelte}[2]{}
    \newcommand{\mycomment}[3]{}
\else
    \newcommand{\mydelete}[2]{\color{#1}{\sout{#2}}}  
    \newcommand{\mycomment}[3]{\color{#1}{[\small \textbf{{#2}:} \textit{#3}]}}  
\fi

\newcommand{\opbr}[0]{{\hat{\mathbf{r}}}}

\newcommand{\rmW}[0]{{\mathrm{(W)}}}
\newcommand{\rmH}[0]{{\mathrm{(H)}}}
\newcommand{\rmSFD}[0]{{\mathrm{S\text{-}FD}}}
\newcommand{\rmMV}[0]{{\mathrm{M\&V}}}
\newcommand{\rmSS}[0]{{\mathrm{S\&S}}}
\newcommand{\rmTEFD}[0]{{\mathrm{TEFD}}}

\definecolor{softgreen}{RGB}{90, 150, 90}

\begin{document}

\title{Accurate calculation of Wannier centers, position matrix, and composite operators
using translationally equivariant and higher-order finite differences}

\author{Jae-Mo Lihm\,\orcidlink{0000-0003-0900-0405}}
\email{jaemo.lihm@gmail.com}
\thanks{These authors contributed equally.}
\affiliation{Department of Physics and Astronomy, Seoul National University, Seoul 08826, Korea}
\affiliation{Center for Theoretical Physics, Seoul National University, Seoul 08826, Korea}
\affiliation{European Theoretical Spectroscopy Facility and Institute of Condensed Matter and Nanosciences, Universit\'e catholique de Louvain, Chemin des \'Etoiles 8, B-1348 Louvain-la-Neuve, Belgium}
\affiliation{Center for Computational Quantum Physics, Flatiron Institute, 162 5th Avenue, New York, NY 10010, USA}
\author{Minsu Ghim\,\orcidlink{0000-0003-0347-735X}}
\email{minsu.ghim.physics@gmail.com}
\thanks{These authors contributed equally.}
\affiliation{Department of Physics and Astronomy, Seoul National University, Seoul 08826, Korea}
\affiliation{Center for Theoretical Physics, Seoul National University, Seoul 08826, Korea}
\author{Seung-Ju Hong\,\orcidlink{0000-0002-9216-8657}}
\email{sjhong6230@snu.ac.kr}
\thanks{These authors contributed equally.}
\affiliation{Department of Physics and Astronomy, Seoul National University, Seoul 08826, Korea}
\affiliation{Center for Theoretical Physics, Seoul National University, Seoul 08826, Korea}
\author{Cheol-Hwan Park\,\orcidlink{0000-0003-1584-6896}}
\email{cheolhwan@snu.ac.kr}
\affiliation{Department of Physics and Astronomy, Seoul National University, Seoul 08826, Korea}
\affiliation{Center for Theoretical Physics, Seoul National University, Seoul 08826, Korea}

\date{April 24, 2026}

\begin{abstract}
The momentum-space derivatives of Bloch wavefunctions are essential for studying quantum geometry and the equilibrium and response properties of solids.
In practical first-principles calculations, these derivatives are obtained via Wannier interpolation of position and related composite matrices.
These matrices are initially evaluated on a coarse $\bk$-point grid using finite-difference approximations and then interpolated to a dense grid.
The accuracy of the finite-difference approximation directly impacts the convergence and reliability of the result.
In this work, we present two key improvements to the finite-difference calculation of position and composite operators for Wannier interpolation.
First, we formulate a translationally equivariant scheme that preserves the underlying symmetries of the system and significantly reduces finite-difference errors.
Second, we introduce a higher-order finite-difference approach that yields a more accurate approximation of the $\bk$-space derivatives by systematically increasing the convergence rate.
From a real-space perspective, these improvements correspond to better approximations of the position operator at the locations of the Wannier functions.
We also present a generalization of the finite-difference scheme, which may reduce the number of finite-difference points while maintaining accuracy.
We demonstrate the effectiveness of our methods by applying them to the calculation of Wannier centers and spreads, electric polarization, off-diagonal position matrix elements, orbital magnetization, and spin Hall conductivity.
Our results demonstrate significant reductions in finite-difference errors, elimination of symmetry-violating errors, and improved convergence with respect to $\bk$-point sampling.
These methods have been implemented in the open-source packages \textsc{Wannier90} and \textsc{WannierBerri}, and they can be readily adopted in other Wannier-based codes with minimal computational overhead.
Our work provides a robust and accurate approach for calculating $\bk$-space derivatives using Wannier interpolation.
\end{abstract}

\maketitle

\section{Introduction}

The reciprocal-space derivatives of the Bloch wavefunctions play a central role in the study of periodic systems.
They are key to studying quantum geometry, which is gaining increasing attention in condensed matter physics~\cite{Xiao2010, Liu2024Geometry, Yu2025Geometry}.
The modern theory of polarization~\cite{1993KingSmithModernPol, 1993RestaModernPol, Vanderbilt1993, 1994OrtizPol, 1994RestaRMP} and orbital magnetization~\cite{2005ThonhauserModernOrb, 2005XiaoModernOrb,2006CeresoliModernOrb,2007ShiModernOrb} has revealed the fundamental role of the Berry phase in describing the equilibrium properties of bulk systems.
The Berry phase and its generalizations serve as essential building blocks for understanding various electrical and optical response properties, including
the anomalous Hall effect~\cite{1996ChangBerryAHC, 1999SundaramBerryAHC,2002OnondaTopoAHC,2002JungwirthAHC},
the bulk photovoltaic effect~\cite{2000SipePhotocurrent},
and the spin Hall effect~\cite{Guo2005SHC,Yao2005SHC}.

In \textit{ab initio} calculations, the $\bk$ derivatives are typically computed using Wannier interpolation, which utilizes a localized real-space representation of the Bloch wavefunctions in terms of the Wannier functions (WFs)~\cite{1937Wannier, 1959KohnBloch, 1997Marzari, 2001Souza, 2012MarzariRMP}.
In real space, the $\bk$-derivative corresponds to the position operator.
The Berry connection can be computed from the Fourier transform of the position matrix elements between WFs, and other geometric quantities follow similarly from the Fourier transform of composite operators involving the position operator.
The implementation of this Wannier interpolation scheme in the \textsc{Wannier90} code~\cite{2008MostofiWannier90, 2020PizziWannier90} has fostered a rich ecosystem of software packages~\cite{Ponce2016EPW, 2018WuWannierTools, tsirkin2021WB, Lee2023EPW, 2024MarrazzoWannier}.

\begin{figure*}[tb]
\centering
\includegraphics[width=0.99\linewidth]{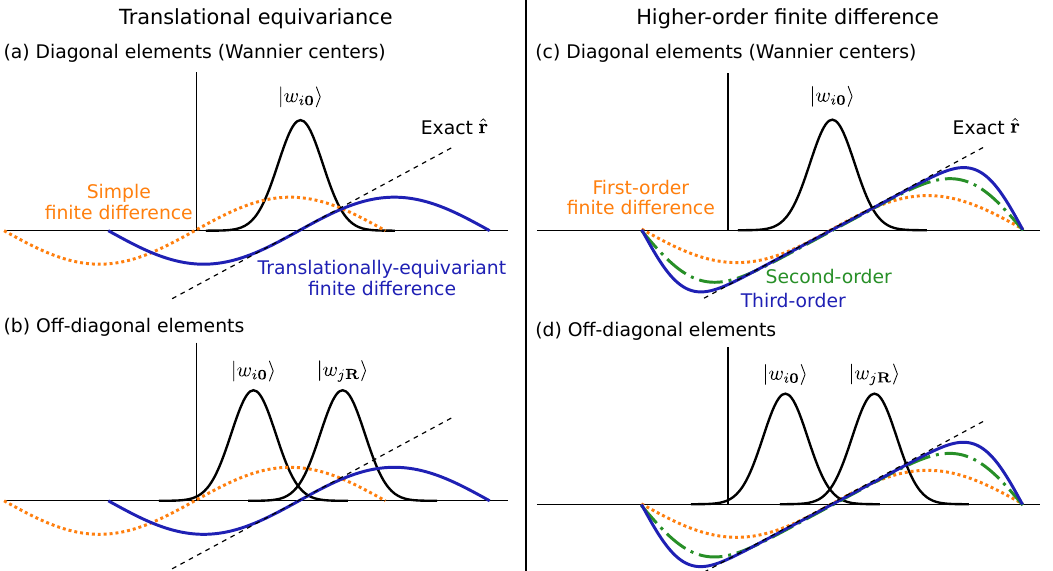}
\caption{
    (a, b) Schematic illustration of the simple finite difference (S-FD) and translationally-equivariant finite difference (TEFD) methods for (a) the diagonal position matrix elements (Wannier centers) and (b) the off-diagonal position matrix elements.
    In the S-FD method, the exact position operator (dashed black lines) is approximated by a sinusoidal function centered at the origin (dotted orange curves).
    In the TEFD method, the sinusoidal approximant is shifted to be centered at the Wannier center or the midpoint between two Wannier centers (solid blue curves), making the FD approximation most accurate at positions where the WFs have the largest magnitude.
    (c, d) Schematic illustration of the higher-order finite difference (HOFD) method.
    As the HOFD order increases, the approximation approaches a sawtooth function, providing a more accurate representation of the exact position operator in the region where the WFs have large magnitudes.
}
\label{fig:schematic}
\end{figure*}

Numerically evaluating the position matrix elements requires a finite difference (FD) approximation of the $\bk$-derivative~\cite{1997Marzari}, as the Brillouin zone can only be sampled discretely.
The accuracy of the FD approximation directly affects the precision of the computed physical quantities.
In principle, one can achieve convergence by using a very dense $\bk$-point grid.
However, the convergence is often slow, as the FD error converges only quadratically with the Brillouin zone sampling.
Therefore, it is desirable to minimize FD errors by employing improved versions of the FD approximation.
While there has been much interest in developing accurate FD approximations for the Wannier centers and electronic polarization~\cite{1997Marzari, SanchezPortal2000Siesta, Soler2002Siesta, 2006Stengel}, little is known about the off-diagonal position matrix elements and other operators.

In this paper, we develop methods that significantly improve the accuracy of the FD calculation for Wannier centers and spreads, position matrix elements, and composite operators involving the position operator.
We present a twofold enhancement to the FD scheme.
First, in Secs.~\ref{sec:position} and \ref{sec:composite}, we highlight the importance of translational equivariance in FD approximations and propose translationally equivariant FD (TEFD) formulas for the position and related matrix elements, respectively.
Second, in Sec.~\ref{sec:higher_order}, we introduce a higher-order FD (HOFD) scheme for $\bk$-space derivatives, which accelerates the convergence of the FD approximation.
These methods are based on a real-space interpretation of the FD approximation, illustrated in \Fig{fig:schematic}.
We demonstrate the effectiveness of the TEFD and HOFD methods by computing Wannier centers and spreads, position matrix elements, electric polarization, orbital magnetization, and spin Hall conductivity.
Our results show that FD errors are significantly reduced: the TEFD method lowers the constant factor in the error and yields results consistent with the crystal symmetry, and the HOFD method further reduces error by improving the exponent of the convergence rate.
In Sec.~\ref{sec:alternative_fd}, we present a more general parametrization of FD approximation using vector-valued coefficients that can be easily combined with the TEFD and HOFD methods.
We conclude in Sec.~\ref{sec:conclusion}.
Table~\ref{tab:summary} summarizes the main FD formulas developed in this paper.

We have implemented these methods in the open-source packages \textsc{Wannier90}~\cite{2008MostofiWannier90, 2020PizziWannier90}, and \textsc{WannierBerri}~\cite{tsirkin2021WB} code.
The developed methods can also be easily incorporated into other software packages, as already have been done for the EPW code~\cite{Lee2023EPW}, to enhance the accuracy of Wannier interpolation schemes for $\bk$-space derivatives.

\section{Translationally equivariant position matrix} \label{sec:position}

\begin{table*}[tb]
\centering
\caption{
    Summary of the key equations.
    Each column lists the equations needed for the calculation of quantities using the simple FD (S-FD), translationally equivariant FD (TEFD), higher-order FD (HOFD), and vector-coefficient FD methods.
    $B[\hat{O}]$ and $C[\hat{O}]$ denote the matrix elements of $\hat{O} (\hat{r} - \bR)$ and $\hat{r} \hat{O} (\hat{r} - \bR)$, respectively [\Eq{eq:real_space_def_BC}].
    For the Wannier interpolation of orbital magnetization, one needs $B[\hat{H}]$ and $C[\hat{H}]$~\cite{2012LopezOrbmag}.
    For spin-Hall conductivity, one needs $B[\hat{s}^\gamma]$ and $B[\hat{s}^\gamma \hat{H}]$~\cite{2018QiaoSHC, 2019RyooSHC}, where $\hat{s}^\gamma$ is the $\gamma$-th component of the spin operator.
}
\label{tab:summary}
{
\renewcommand{\arraystretch}{1.2}
\begin{tabular}{c|c|c|c|c}
\hline\hline
Quantity & S-FD & TEFD & HOFD & Vector-coefficient FD\\
\hline
Wannier centers $\br_i$ & \Eq{eq:center_FD} & \makecell{\Eq{eq:tr_r_fd_self_consistent}} &
\multirow{4}{*}{\shortstack{
    \Eqs{eq:fd_nearest_nd}, \eqref{eq:hofd_multiples_b}, \eqref{eq:cramers_nd_}
}} &
\multirow{4}{*}{\shortstack{
    FD and HOFD: \Eqs{eq:Bx_eq_c_sol}, \eqref{eq:fd_general}, \eqref{eq:new_method_qb_vector}\\
    TEFD: Replacing $c_{\abs{\bb}} \bb$ with $\mb{g}_\bb$ [Eq.~\eqref{eq:new_method_qb_vector}]
}}
\\ \cline{1-3}
Position matrix $\br_{\ijR}$ & \Eq{eq:r_fd} & \Eq{eq:tr_r_fd_trinv} &\\
\cline{1-3}
\makecell[l]{Composite $B_{\ijR}^\alpha[\hat{O}]$} & \Eq{eq:BB_fd} & \Eq{eq:BB_trinv} &\\
\cline{1-3}
\makecell[l]{Composite $C_{\ijR}^{\alpha\beta}[\hat{O}]$} & \Eq{eq:CC_fd} & \Eq{eq:CC_trinv} &\\
\hline\hline
\end{tabular}
}
\end{table*}

\subsection{Wannier functions and position matrix elements} \label{sec:wannier}

We begin by reviewing the definitions of WFs and the FD expression for position matrix elements.
The Wannier and Bloch functions are related by a Fourier transformation that includes a $\bk$-dependent gauge matrix $U_{mi\bk}$:
\begin{equation} \label{eq:wf_def}
    \ket{w_\iR} = \frac{1}{\sqrt{N_\bk}} \sum_{\mk} \ket{\psi_\mk} U_{mi\bk} e^{-i\bk\cdot\bR} \,,
\end{equation}
and
\begin{equation} \label{eq:w_to_psi}
    \ket{\psi_\mk} = \frac{1}{\sqrt{N_\bk}} \sum_{\iR} \ket{w_\iR} U_{im\bk}^\dagger \expikR \,.
\end{equation}
Here, $\ket{\psi_\mk}$ is the Bloch eigenstate for band $m$ at wavevector $\bk$,
$\ket{w_\iR}$ is the WF with orbital index $i$ localized at unit cell $\bR$,
and $N_\bk$ is the number of $\bk$ points sampled in the Brillouin zone, corresponding to the size of the Born-von Karman supercell.
We denote a Bloch wavefunction in the Wannier gauge as
\begin{align} \label{eq:psi_ik_def}
    \ket{\psi^\rmW_\ik}
    &= \sum_m \ket{\psi_\mk} U_{mi\bk}
    \nnnl
    &= \frac{1}{\sqrt{N_\bk}} \sum_{\iR} \ket{w_\iR} \expikR \,,
\end{align}
and its periodic part as
\begin{equation} \label{eq:u_ik_def}
    \ket{u^\rmW_\ik}
    = e^{-i\bk\cdot\opbr}\ket{\psi^\rmW_\ik} \,,
\end{equation}
with $\opbr = i\nabla_\bk$ the position operator.
We reserve $m$ and $n$ for band indices, and $i$ and $j$ for WF indices.
The matrix element of the position operator between the WFs is defined as~\cite{Blount1962}
\begin{align} \label{eq:r_integral}
    \br_{\ijR}
    &= \mel{w_{i\mb{0}}}{\opbr}{w_\jR}
    \nnnl
    &= \frac{i}{N_\bk} \sum_\bk \, e^{-i\bk\cdot\bR}
    \braket{u^\rmW_\ik}{\nabla_\bk u^\rmW_\jk} \,.
\end{align}

In practice, both the Bloch eigenstates and the gauge matrices are computed on a discrete grid of $\bk$ points.
Thus, the derivative in \Eq{eq:r_integral} must be approximated using an FD scheme.
The simplest such approximation takes the form
\begin{equation} \label{eq:fd_naive}
    \nabla_\bk f(\bk) \approx \sum_\bb c_{\abs{\bb}} \bb f(\bkb) \,,
\end{equation}
where $\bb$'s are the FD vectors connecting neighboring grid points, and $c_{\abs{\bb}}$'s are the associated FD weights.
These weights are chosen to satisfy the condition~\cite{1997Marzari}
\begin{equation} \label{eq:c_b_sum}
    \sum_\bb c_{\abs{\bb}} b^\alpha b^\beta = \delta_{\alpha\beta} \,.
\end{equation}
For a cubic lattice, the $\bb$ vectors can be the six vectors that connect one $\bk$ grid point to its nearest neighbors.
For general lattices, one determines the smallest cutoff distance $b_{\rm cut}$ such that including some of $\bb$ vectors with $\abs{\bb} \leq b_{\rm cut}$ allows for a solution to \Eq{eq:c_b_sum}.
The selection of $\bb$ vectors will be discussed in more detail in \Sec{sec:higher_order} and \Sec{sec:alternative_fd}.

In real space, the FD approximation of \Eq{eq:fd_naive} corresponds to approximating the position operator with a supercell-periodic operator
\begin{equation} \label{eq:opr_fd}
    \opbr
    \approx \opbr^{\rmSFD}
    = i \sum_{\bb} c_{\abs{\bb}} \bb \, e^{-i\bb \cdot \opbr}\,.
\end{equation}
We call this the simple FD (S-FD) approximation.
Using this approximation to rewrite the derivative in \Eq{eq:r_integral}, we find~\cite{1997Marzari}
\begin{align} \label{eq:r_fd}
    \br_\ijR^{\rmSFD}
    &= \mel{w_{i\mb{0}}}{\opbr^{\rmSFD}}{w_\jR}
    \nnnl
    &= \frac{i}{N_\bk} \sum_{\bk\bb} c_{\abs{\bb}} \bb \, e^{-i\bk\cdot\bR} \,
    \mel{\psi^\rmW_\ik}{e^{-i\bb\cdot\opbr}}{\psi^\rmW_\jkb}
    \nnnl
    &= \frac{i}{N_\bk} \sum_{\bk\bb} c_{\abs{\bb}} \bb \, e^{-i\bk\cdot\bR} \, M_{ij}^\kb \,,
\end{align}
where
\begin{align} \label{eq:M_def}
    M_{ij}^\kb
    &= \braket{u^\rmW_\ik}{u^\rmW_\jkb}
    \nnnl
    &= \sum_{mn} U^\dagger_{im\bk} \braket{u_\mk}{u_\nkb} \, U_{nj\bkb}
\end{align}
is the overlap matrix.
This matrix can also be expressed in terms of the WFs as
\begin{align} \label{eq:M_ijR}
    M_{ij}^\kb
    &= \frac{1}{N_\bk} \sum_{\bR \bRp} e^{-i\bk\cdot\bR}
    \mel{w_{\iR}}{e^{-i\bb\cdot\opbr}}{w_{\jRp}}
    e^{i(\bk + \bb)\cdot\bRp} \nnnl
    &= \sum_{\bR} \expval{e^{-i\bb\cdot\opbr}}_{\ijR} e^{i(\bk + \bb)\cdot\bR} \,,
\end{align}
where we introduced the shorthand notation,
\begin{equation}
    \expval{A}_\ijR \equiv \mel{w_{i\mb{0}}}{A}{w_{j\bR}} \,.
\end{equation}
From \Eq{eq:M_ijR}, it also follows that
\begin{equation} \label{eq:M_ii}
    \frac{1}{N_\bk} \sum_\bk M_{ii}^\kb = \expval{e^{-i\bb\cdot\opbr}}_i \,,
\end{equation}
where $\expval{A}_i \equiv \mel{w_{i\mb{0}}}{A}{w_{i\mb{0}}}$.

The diagonal elements of the position matrix correspond to the Wannier centers, defined as
\begin{equation}
    \br_{i} \equiv \br_{ii\mb{0}} = \mel{w_{i\mb{0}}}{\opbr}{w_{i\mb{0}}} \,.
\end{equation}
The S-FD approximation of Wannier centers can be obtained by taking the diagonal part of \Eq{eq:r_fd}, yielding
\begin{equation} \label{eq:center_FD}
    \br_i^{\rmSFD} = \frac{i}{N_\bk} \sum_{\bk\bb} c_{\abs{\bb}} \bb \, M_{ii}^\kb \,.
\end{equation}
(The $-1$ term in Eq.~(22) of Ref.~\cite{1997Marzari} vanishes after the $\bb$ sum due to cancellation between $\bb$ and $-\bb$.)
In practice, different expressions are used to calculate Wannier centers, as discussed in \Sec{sec:tr_center}.

\subsection{Translational equivariance} \label{sec:tr_eqvar}

Although \Eqs{eq:r_fd} and \eqref{eq:center_FD} are valid FD approximations that converge to the exact result in the limit of an infinitely dense grid, it has been shown that they do not transform correctly under a uniform translation of the entire system~\cite{1997Marzari}.
In this section, we define this concept of translational equivariance.

Consider a uniform translation of the entire system, including atoms, electrons, and WFs, by a fixed, arbitrary displacement $\bd$, which does not have to be a lattice vector.
According to its definition, $\br_{\ijR} = \mel{w_{i\mb{0}}}{\opbr}{w_\jR}$, the position matrix elements should transform as
\begin{equation} \label{eq:tr_r_transform}
    \br'_\ijR = \br_\ijR + \bd \, \delta_{ij} \delta_{\bR\mb{0}} \,,
\end{equation}
where the quantities after translation are denoted with a prime.
In effect, only $\opbr$ is changed by $\opbr+\bd$, while $\bR$ remains fixed.
The diagonal elements, which correspond to the Wannier centers, shift by $\bd$, while the off-diagonal elements remain unchanged due to the orthogonality of the WFs.
Similar transformation rules apply to composite operators as well [see \Eqs{eq:real_space_def_BC} and \eqref{eq:transf_BC}].
We call an approximate expression \textit{translationally equivariant} (or simply \textit{equivariant}) if it yields values that satisfy the corresponding transformation rule, even in the presence of FD errors.
For example, an FD expression for the position matrix is equivariant if the resulting matrix elements transform according to \Eq{eq:tr_r_transform}.

In the following, we derive the transformation properties of various FD expressions.
To this end, we will use the fact that the Bloch wavefunctions and WFs transform as
\begin{equation}
    \psi'_\mk(\br) = \psi_\mk(\br - \bd) \,,\ w'_\iR(\br) = w_\iR(\br - \bd) \,.
\end{equation}
Since the momentum operator $\opmbp / \hbar = -i \nabla_\br$ generates translations, the translated wavefunctions can be written as
\begin{equation} \label{eq:wave_transform}
    \ket{\psi^{\rmW \prime}_\ik} = e^{-i\frac{\opmbp}{\hbar}\cdot\bd} \ket{\psi_\ik^{\rmW}} \,,\ %
    \ket{u^{\rmW \prime}_\ik} = e^{-i(\frac{\opmbp}{\hbar} + \bk)\cdot\bd} \ket{u^{\rmW}_\ik} \,,
\end{equation}
the latter of which follows from $\ket{u^{\rmW \prime}_\ik}=e^{-i\bk\cdot\opbr}\ket{\psi^{\rmW \prime}_\ik}$.
The transformation of the overlap matrix [\Eq{eq:M_def}] then follows as
\begin{align} \label{eq:tr_M_transform}
    M_{ij}^{\prime\,\kb}
    = \braket{u^{\rmW \prime}_\ik}{u^{\rmW \prime}_\jkb}
    &= e^{-i\bb \cdot\bd} \braket{u^\rmW_\ik}{u^\rmW_\jkb}
    \nnnl
    &= e^{-i\bb\cdot\bd} M_{ij}^\kb \,.
\end{align}
In the second equality, we used \Eq{eq:wave_transform}.

\subsection{Wannier centers} \label{sec:tr_center}

Let us first focus on the Wannier centers.
It was shown in Ref.~\cite{1997Marzari} that the S-FD formula \eqref{eq:center_FD} is not equivariant:
\begin{equation} \label{eq:tr_center_FD_prime}
    \br_i^{\prime\,\rmSFD}
        = \frac{i}{N_\bk} \sum_{\bk\bb} c_{\abs{\bb}} \bb \, e^{-i \bb \cdot \bd} M_{ii}^\kb
    \neq \br_i^{\rmSFD} + \bd \,,
\end{equation}
which we can easily show by putting \Eq{eq:tr_M_transform} into \Eq{eq:center_FD}.
To better understand how the equivariance is broken, we expand the FD expression in powers of $b$ and analyze the leading FD error.
Using \Eq{eq:M_ii}, we obtain
\begin{align} \label{eq:tr_center_fd_expand}
    \br_{i}^{\rmSFD}
    &= i \sum_{\bb} c_{\abs{\bb}} \bb \, \expval{e^{-i\bb\cdot\opbr}}_i \nnnl
    &= i \sum_{\bb} c_{\abs{\bb}} \bb \, \Bigl( -i \expval{\bb\cdot\opbr}_i + \frac{(-i)^3}{6} \expval{(\bb \cdot \opbr)^3}_i + O(b^5) \Bigr) \nnnl
    &= \br_i
    - \frac{1}{6} \sum_{\bb} c_{\abs{\bb}} \bb \, \expval{(\bb \cdot \opbr)^3}_i
    + O(b^4)
\end{align}
Note that terms involving an odd number of $\bb$ vectors vanish because $c_{\abs{\bb}} = c_{\abs{-\bb}}$, and that $c_b$ is of $O(b^{-2})$, making the second term of order $O(b^2)$.
The leading FD error is proportional to $\expval{\opbr^3}_i$, which is not equivariant.
Under a uniform translation, it transforms as
\begin{equation}
    \expval{\opbr^3}'_i
    = \expval{(\opbr + \bd)^3}_i
    \neq \expval{\opbr^3}_i \,.
\end{equation}

Moreover, this FD error is not necessarily small even when the WFs are well localized around their centers.
Defining
\begin{equation}
    \delta_i \opbr = \opbr - \br_i \,,
\end{equation}
we find that the FD error is roughly proportional to
\begin{align} \label{eq:center_fd_error}
    \expval{\opbr^3}_i
    &= \expval{(\delta_i \opbr + \br_i)^3}_i
    \nnnl
    &= \expval{(\delta_i \opbr)^3}_i
    + 3 \expval{(\delta_i \opbr)^2}_i \, \br_i
    + 3 \expval{\delta_i \opbr}_i (\br_i)^2
    + (\br_i)^3 \,.
\end{align}
The localization of the WFs ensures only that $\expval{(\delta_i \opbr)^n}_\ijR$ is small; it does not imply that $\br_i$ itself is small.
As a result, the FD error can become large when the Wannier center $\br_i$ is far from the origin.

The fundamental issue with the S-FD approximation is that \Eq{eq:opr_fd} is valid only near the origin~\cite{2006Stengel}.
To accurately calculate the Wannier centers, the position operator must be approximated accurately around $\br_i$, which may lie far from the origin, where the S-FD approximation does not work well.
As a result, the accuracy of the S-FD expression deteriorates as the WF center moves away from the origin.
This behavior is illustrated by the schematic result for S-FD in \Fig{fig:schematic}(a).

To address the problem of broken equivariance, Marzari and Vanderbilt (M\&V) proposed an alternative FD expression~\cite{1997Marzari}, given by
\begin{equation} \label{eq:center_MV}
    \br_{i}^{\rmMV} = -\frac{1}{N_\bk} \sum_{\bk\bb} c_{\abs{\bb}} \bb \Im \ln M_{ii}^\kb \,.
\end{equation}
This formula is equivariant:
\begin{align} \label{eq:tr_center_MD_prime}
    \br_{i}^{\prime\,\rmMV}
    &= -\frac{1}{N_\bk} \sum_{\bk\bb} c_{\abs{\bb}} \bb \Im \ln (e^{-i\bb\cdot\bd} M_{ii}^\kb)
    \nnnl
    &= \br_{i}^{\rmMV} + \frac{1}{N_\bk} \sum_{\bk\bb} c_{\abs{\bb}} \bb \, (\bb \cdot \bd)
    \nnnl
    &= \br_{i}^{\rmMV} + \bd \,,
\end{align}
where we used \Eq{eq:tr_M_transform} in the first equality and \Eq{eq:c_b_sum} in the last equality.
A related variant was later introduced by Stengel and Spaldin (S\&S)~\cite{2006Stengel}, in which the Brillouin zone average is taken inside the logarithm:
\begin{equation} \label{eq:center_SS}
    \br_{i}^{\rmSS} = - \sum_{\bb} c_{\abs{\bb}} \bb \Im \ln \Bigl( \frac{1}{N_\bk} \sum_\bk M_{ii}^\kb \Bigr) \,.
\end{equation}
This formula is also equivariant:
\begin{align} \label{eq:tr_center_SD_prime}
    \br_{i}^{\prime\,\rmSS}
    &= - \sum_{\bb} c_{\abs{\bb}} \bb \Im \ln \Bigl[ e^{-i\bb\cdot\bd} \Bigl( \frac{1}{N_\bk} \sum_\bk M_{ii}^\kb \Bigr) \Bigr]
    \nnnl
    &= \br_{i}^{\rmSS} + \bd \,.
\end{align}

We now examine the leading FD error in these equivariant expressions.
For the M\&V formula, we begin by expanding the overlap matrix [\Eq{eq:M_ijR}] as
\begin{align}
    &M_{ij}^\kb
    = \sum_{\bR} \expval{e^{-i\bb\cdot(\opbr-\bR)}}_{\ijR} e^{i\bk\cdot\bR}
    \nnnl
    &= \sum_{\bR} \Bigl[ \braket{w_{i\mb{0}}}{w_\jR}
    + \sum_{n=1}^\infty \frac{(-i)^n}{n!} \expval{(\bb \cdot (\opbr-\bR))^n}_\ijR \Bigr] e^{i\bk\cdot\bR}
    \nnnl
    &= \delta_{ij}
    + \sum_{\bR} \sum_{n=1}^\infty \frac{(-i)^n}{n!} \expval{(\bb \cdot (\opbr-\bR))^n}_\ijR \, e^{i\bk\cdot\bR}
    \,.
\end{align}
Then, using the expansion
\begin{equation}
    \ln (1+x) = x - \frac{x^2}{2} + \frac{x^3}{3} + O(x^4) \,,
\end{equation}
and defining
\begin{equation}
    p^{(\bk,\bb)}_{n, i} = \sum_{\bR} \expval{(\bb \cdot (\opbr - \bR))^{n}}_\iiR e^{i\bk\cdot\bR} \,,
\end{equation}
we find that the Taylor expansion of $\ln M_{ii}^\kb$ reads
\begin{align}
    \ln M_{ii}^\kb
    &= -i p^{(\bk,\bb)}_{1, i}
    + \bb^2 (\cdots)
    \nnnl
    &+ \frac{i}{6} \Bigl[ p^{(\bk,\bb)}_{3, i} - 3 p^{(\bk,\bb)}_{1, i} p^{(\bk,\bb)}_{2, i}
    + 2 \bigl( p^{(\bk,\bb)}_{1, i} \bigr)^3 \Bigr]
    \nnnl
    &+ \bb^4 (\cdots) + O(b^5)
    \,.
\end{align}
The term $p^{(\bk,\bb)}_{n, i}$ for $n = 1, 2, 3$ can be expanded further as
\begin{align}
    p^{(\bk,\bb)}_{1, i}
    &= \sum_{\bR} e^{i\bk\cdot\bR} \expval{\bb \cdot \delta_i \opbr}_\iiR
    + \bb \cdot \br_i \,,
    \\
    p^{(\bk,\bb)}_{2, i}
    &= \sum_{\bR} e^{i\bk\cdot\bR} \Bigl[
        \expval{(\bb \cdot \delta_i \opbr)^2}_\iiR
    \nnnl
        &\quad + 2 \expval{\bb \cdot \delta_i \opbr}_\iiR (\bb \cdot (\br_i - \bR))
    \Bigr]
    + (\bb \cdot \br_i)^2 \,,
    \\
    p^{(\bk,\bb)}_{3, i}
    &= \sum_{\bR} e^{i\bk\cdot\bR} \Bigl[
        \expval{(\bb \cdot \delta_i \opbr)^3}_\iiR
    \nnnl
        &\quad + 3 \expval{(\bb \cdot \delta_i \opbr)^2}_\iiR (\bb \cdot (\br_i - \bR))
    \nnnl
        &\quad + 3 \expval{\bb \cdot \delta_i \opbr}_\iiR (\bb \cdot (\br_i - \bR))^2
    \Bigr]
    + (\bb \cdot \br_i)^3
    \,.
\end{align}
Here, we used the orthogonality of the WFs that ensures $\expval{c}_\iiR = c \, \delta_{\bR\mb{0}}$ for any constant $c$.
Substituting these expressions into \Eq{eq:center_MV}, we obtain the Taylor expansion of the M\&V formula, which reads
\begin{align} \label{eq:center_MV_expand}
    &\br_{i}^{\rmMV} - \br_i
    \nnnl
    &= - \frac{1}{6} \sum_{\bb} c_{\abs{\bb}} \bb \Bigg[
    \expval{(\bb \cdot \delta_i \opbr)^3}_i
    -3 \sum_{\bR}
    \expval{\bb \cdot \delta_i \opbr}_\iiR
    \expval{(\bb \cdot \delta_i \opbr)^2}_{ii\bmR}
    \nnnl
    &\ + 2 \sum_{\bR \bRp}
    \expval{\bb \cdot \delta_i \opbr}_\iiR
    \expval{\bb \cdot \delta_i \opbr}_{ii\bRp}
    \expval{\bb \cdot \delta_i \opbr}_{ii\bmR\mb{-R'}}
    \Bigg]
    + O(b^4) \,.
\end{align}
We used $\expval{\bb \cdot \delta_i \opbr}_i = 0$ to eliminate some terms.

For the S\&S formula, we substitute \Eq{eq:M_ii} into \Eq{eq:center_SS} to find
\begin{align} \label{eq:center_SS_expand}
    &\br_{i}^{\rmSS}\nnnl
    &= - \sum_{\bb} c_{\abs{\bb}} \bb \, \Im \ln \expval{e^{-i\bb\cdot\opbr}}_i \nnnl
    &= - \sum_{\bb} c_{\abs{\bb}} \bb \, \Im  \Big( -i \expval{\bb\cdot\opbr}_i
    + \frac{i}{6} \expval{(\bb \cdot \delta_i \opbr)^3}_i + O(b^5) \Big)
    \nnnl
    &= \br_i
    - \frac{1}{6} \sum_{\bb} c_{\abs{\bb}} \bb \, \expval{(\bb \cdot \delta_i \opbr)^3}_i
    + O(b^4) \,.
\end{align}
In the second equality, we used the cumulant series expansion
\begin{multline}
    \ln \expval{e^{-i\bb\cdot\opbr}}_i
    = -i \expval{\bb\cdot\opbr}_i
    + \frac{(-i)^2}{2} \expval{(\bb\cdot \delta_i \opbr)^2}_i
    \\
    + \frac{(-i)^3}{6} \expval{(\bb\cdot \delta_i \opbr)^3}_i
    + O(b^4) \,.
\end{multline}

In \Eqs{eq:center_MV_expand} and \eqref{eq:center_SS_expand}, the leading FD error depends only on the matrix elements of $b^{-1}(b\,\delta_i \opbr)^3$, which are invariant under translation.
Also, if the WF is localized around its center $\br_i$ with a width $\sigma_i$, the expectation value of $b^{-1}(b\,\delta_i \opbr)^3$ is bounded from above by a term of order $b^2\,\sigma_i^3$.
Thus, the M\&V and S\&S formulas exhibit small errors if the WFs are well localized, offering an improvement over the S-FD formula.

In passing, we note that the additional terms in \Eq{eq:center_MV_expand} compared to \Eq{eq:center_SS_expand} account for the lack of size-consistency in the M\&V formula: using a unit cell with $\bk$ point sampling and a supercell with $\Gamma$-point sampling does not yield the same result for the Wannier centers in the M\&V formula.
Both the S\&S method and the method we propose in the next section are size-consistent: see \Sec{sec:size_consistency} for a proof.

\subsection{Translationally equivariant position matrix}

The M\&V and S\&S formulas for the Wannier centers are not directly applicable to the calculation of the full position matrix.
An equivariant FD formula for the full position matrix has not yet been developed.
Hence, the S-FD formula [\Eq{eq:r_fd}], implemented in the Wannier90 code~\cite{2008MostofiWannier90, 2020PizziWannier90}, is widely used.

However, for the Wannier centers, the off-diagonal position matrix elements from the S-FD formula also suffer from the same issue of broken equivariance.
Taylor expanding \Eq{eq:r_fd} yields
\begin{align} \label{eq:tr_mel_fd_expand}
    \br_\ijR^\rmSFD
    &= i \sum_{\bb} c_{\abs{\bb}} \bb \mel{w_{i\mb{0}}}{e^{-i\bb\cdot\opbr}}{w_{j\bR}}
    \nnnl &= \expval{\opbr}_\ijR
    - \frac{1}{6} \sum_{\bb} c_{\abs{\bb}} \bb \expval{(\bb \cdot \opbr)^3}_\ijR
    + O(b^4) \,,
\end{align}
where the leading error term is proportional to $\expval{\opbr^3}_\ijR$.
We define a position operator centered at the midpoint of the two WFs:
\begin{equation}
    \delta_\ijR \, \opbr = \opbr - \bar{\br}_{\ijR} \,,
\end{equation}
where
\begin{equation} \label{eq:tr_r_midpoint_def}
    \bar{\br}_{\ijR} = \frac{\br_i + \br_j + \bR}{2} \,.
\end{equation}
For the off-diagonal element $(i, \mb{0}) \neq (j, \mb{R})$, the FD error is roughly proportional to
\begin{align} \label{eq:tr_mel_fd_error}
    \expval{\opbr^3}_\ijR
    &= \expval{(\delta_\ijR \, \opbr + \bar{\br}_{\ijR})^3}_\ijR
    \nnnl
    &= \expval{(\delta_\ijR \, \opbr)^3}_\ijR
    + 3 \expval{(\delta_\ijR \, \opbr)^2}_\ijR \bar{\br}_{\ijR}
    \nnnl
    &\quad + 3 \expval{\delta_\ijR \, \opbr}_\ijR (\bar{\br}_{\ijR})^2 \,.
\end{align}
Compared to the diagonal case [\Eq{eq:center_fd_error}], the last term $(\bar{\br}_{\ijR})^3$, corresponding to $(\br_i)^3$, is absent due to the orthogonality of the WFs.
Nevertheless, the FD error is not translationally invariant and can be large whenever the midpoint $\bar{\br}_{\ijR}$ is far from the origin, even if all WFs are localized.

We now propose an equivariant formula for the full position matrix.
To ensure equivariance, we adjust the periodic approximant of the position operator $\opbr$ so that it is most accurate at the midpoint of the two WFs, where the product of the WFs typically has the largest magnitude.
This approach leads to the TEFD approximation
\begin{equation} \label{eq:opr_TEFD}
    \opbr
    \approx \opbr^{\rmTEFD}_\ijR
    = \bar{\br}_{\ijR}
    + i \sum_\bb c_{\abs{\bb}} \bb \, e^{-i\bb \cdot (\opbr - \bar{\br}_{\ijR})}
    \,.
\end{equation}
This operator is illustrated in \Fig{fig:schematic}(a, b).
Compared to the S-FD approximant, the TEFD approximant is centered at the midpoint $\bar{\br}_{\ijR}$ of the two WFs, and thus provides a good approximation to the position operator at that point.
A similar idea has been applied in the calculation of electric polarization (i.e., the diagonal elements of the position matrix) with a local combination of atomic orbital basis set~\cite{SanchezPortal2000Siesta, Soler2002Siesta}.

By evaluating the matrix elements of the TEFD operator, we obtain
\begin{align} \label{eq:tr_r_fd_trinv}
    &\quad \br_{\ijR}^\rmTEFD
    \\
    &= \mel{w_{i\mb{0}}}{\opbr^{\rmTEFD}_\ijR}{w_{j\bR}}
    \nnnl
    &= \bar{\br}_{\ijR} \delta_{ij} \delta_{\bR\mb{0}}
    + i \sum_{\bb} c_{\abs{\bb}} \bb \mel{w_{i\mb{0}}}{e^{-i\bb \cdot \left(\opbr- \bar{\br}_{\ijR} \right) }}{w_{j\bR}}
    \nnnl
    &= \br_i \delta_{ij} \delta_{\bR\mb{0}}
    \nnnl
    &+ \frac{i}{N_\bk} \sum_{\bk\bb} c_{\abs{\bb}} \bb \, e^{i\bb \cdot \bar{\br}_{\ijR}} \mel{\psi^{({\textrm W})}_{i\bk}}{e^{-i\bb \cdot \opbr}}{\psi^{({\textrm W})}_{j\mb{k+b}}} e^{-i(\bk+\bb)\cdot\bR}
    \nnnl
    &= \br_i \delta_{ij} \delta_{\bR\mb{0}}
    + \frac{i}{N_\bk} \sum_{\bb} c_{\abs{\bb}} \bb \, e^{i\bb\cdot \bar{\br}_{\ijR}} \sum_{\bk} e^{-i(\bkb)\cdot\bR} M_{ij}^\kb \,.
    \nonumber
\end{align}
By Taylor expanding \Eq{eq:opr_TEFD} as
\begin{align} \label{eq:opr_TEFD_taylor}
    \opbr^{\rmTEFD}_\ijR
    &= \bar{\br}_{\ijR}
    + i \sum_\bb c_{\abs{\bb}} \bb \, (-i\bb \cdot (\opbr - \bar{\br}_{\ijR}))
    \nnnl
    &+ i \sum_\bb c_{\abs{\bb}} \bb \, \frac{(-i)^3}{6} \bigl[ \bb \cdot (\opbr - \bar{\br}_{\ijR}) \bigr]^3
    + O(b^4)
    \nnnl
    &= \opbr
    - \frac{1}{6} \sum_\bb c_{\abs{\bb}} \bb \, \bigl[ \bb \cdot (\opbr - \bar{\br}_{\ijR}) \bigr]^3
    + O(b^4)
    \,,
\end{align}
we find the leading FD error of the TEFD formula:
\begin{align} \label{eq:tr_r_fd_trinv_error}
    &\quad \br_{\ijR}^\rmTEFD - \br_\ijR
    \\
    &=
    - \frac{1}{6} \sum_\bb c_{\abs{\bb}} \bb \,
    \mel{w_{i\mb{0}}}{\bigl[ \bb \cdot (\opbr - \bar{\br}_{\ijR}) \bigr]^3}{w_\jR}
    + O(b^4)
    \,.
    \nonumber
\end{align}
This FD error is translationally invariant and remains small if the WFs are well localized.
\Equ{eq:tr_r_fd_trinv} represents the first central result of this paper: a translationally equivariant FD expression for the full position matrix.
The proposed formula is straightforward to implement with only a negligible increase in computational cost.

To evaluate the TEFD formula \Eq{eq:tr_r_fd_trinv}, the Wannier centers must be known in advance.
One approach is to first compute the Wannier centers using the M\&V or S\&S formulas and then calculate the off-diagonal elements using \Eq{eq:tr_r_fd_trinv}.
Alternatively, the Wannier centers can be computed by solving a self-consistent equation: $\br_i^\rmTEFD$ for the $(n+1)$-th iteration, $\br_{i,(n+1)}^\rmTEFD$, is given by
\begin{equation} \label{eq:tr_r_fd_self_consistent}
    \br_{i,(n+1)}^\rmTEFD
    = \br_{i,(n)}^\rmTEFD
    + \frac{i}{N_\bk} \sum_\bb c_{\abs{\bb}} \bb \, e^{i\bb \cdot \br_{i,(n)}^\rmTEFD}
    \sum_\bk M_{ii}^{(\bk,\bb)} \,,
\end{equation}
which comes from the diagonal part of \Eq{eq:tr_r_fd_trinv}.
A similar self-consistent iteration, using a sawtooth representation of the position operator rather than an FD approximation, was used in Ref.~\cite{2006Stengel}.
In practice, we find that the self-consistent TEFD method produces Wannier centers similar to those obtained with the M\&V and S\&S methods (see \Sec{sec:trinv_results}).

If one considers only the equivariance, there is an ambiguity in determining the translationally equivariant formula.
Instead of the midpoint $\bar{\br}_\ijR = ({\br_i + \br_j + \bR})/{2}$ [\Eq{eq:tr_r_midpoint_def}], one can subtract from the position operator any vector that translates with the system in such a way that
\begin{equation}
    \bar{\br}_\ijR^{(f)} = f \br_i + (1-f) (\br_j + \bR)\,,
\end{equation}
where $f$ is an arbitrary real number.
Note that, under a translation of the entire system by ${\bf d}$, $\br_i\rightarrow\br_i+\textbf{d}$ and $\br_j\rightarrow\br_j+\textbf{d}$, and hence, $\bar{\br}_\ijR^{(f)}\rightarrow\bar{\br}_\ijR^{(f)}+\textbf{d}$, satisfying translational equivariance, independent of $f$.
To fix $f$ to $1/2$, we impose the Hermiticity of the position matrix.
For the off-diagonal components ($i \neq j$ or $\bR \neq \mb{0}$), Hermiticity requires
\begin{align}
    \br_\ijR
    &= \mel{w_{i\mb{0}}}{\opbr}{w_\jR}
    \nnnl
    &= \mel{w_\jR}{\opbr}{w_{i\mb{0}}}^*
    = \mel{w_{j\mb{0}}}{\opbr}{w_{i\bmR}}^*
    = \br_{ji\bmR}^* \,.
\end{align}
To satisfy this condition, the TEFD formula [\Eq{eq:tr_r_fd_trinv}], with $\bar{\br}_\ijR$ replaced by $\bar{\br}_\ijR^{(f)}$, should satisfy
\begin{align} \label{eq:tr_r_herm_der1}
    &\frac{i}{N_\bk} \sum_{\bb} c_{\abs{\bb}} \bb \, e^{i\bb\cdot \bar{\br}^{(f)}_{\ijR}} \sum_{\bk} e^{-i(\bkb)\cdot\bR} M_{ij}^\kb
    \nnnl
    &= \biggl( \frac{i}{N_\bk} \sum_{\bb} c_{\abs{\bb}} \bb \, e^{i\bb\cdot \bar{\br}^{(f)}_{ji\bmR}} \sum_{\bk} e^{i(\bkb)\cdot\bR} M_{ji}^\kb
    \biggr)^*
    \nnnl
    &= -\frac{i}{N_\bk} \sum_{\bb} c_{\abs{\bb}} \bb \, e^{-i\bb\cdot \bar{\br}^{(f)}_{ji\bmR}} \sum_{\bk} e^{-i(\bkb)\cdot\bR} M_{ij}^{(\bkb,-\bb)}
    \nnnl
    &= \frac{i}{N_\bk} \sum_{\bb} c_{\abs{\bb}} \bb \, e^{i\bb\cdot \bar{\br}^{(f)}_{ji\bmR}} \sum_{\bk} e^{-i\bk\cdot\bR} M_{ij}^\kb
    \,.
\end{align}
In the second equality, we used
\begin{equation}
    \bigl( M_{ji}^\kb \bigr)^*
    = M_{ij}^{(\bkb,-\bb)} \,,
\end{equation}
and in the third equality, we changed variables $\bk+\bb$ to $\bk$ and then $\bb$ to $-\bb$.
From \Eq{eq:tr_r_herm_der1}, we find
\begin{align}
    \bar{\br}^{(f)}_{\ijR} - \bR
    &= f \br_i + (1 - f)\br_j - f \bR
    \nnnl
    = \bar{\br}^{(f)}_{ji\bmR}
    &= (1 - f)\br_i + f \br_j - (1 - f) \bR \,.
\end{align}
Thus, we conclude $f = 1 - f$ and $f = 1/2$.

This result implies that the TEFD formula [\Eq{eq:tr_r_fd_trinv}], with $f = 1/2$, is Hermitian even in the presence of FD errors.
In contrast, the S-FD formula [\Eq{eq:r_fd}] breaks Hermiticity:
\begin{align} \label{eq:r_Sfd_herm_problem}
    (\br_{ji-\bR}^{\rmSFD})^*
    &= -\frac{i}{N_\bk} \sum_{\bk\bb} c_{\abs{\bb}} \bb \, e^{-i\bk\cdot\bR} \, \bigl( M_{ji}^\kb \bigr)^*
    \nnnl
    &= -\frac{i}{N_\bk} \sum_{\bk\bb} c_{\abs{\bb}} \bb \, e^{-i\bk\cdot\bR} \, M_{ij}^{(\bk+\bb, -\bb)}
    \nnnl
    &= \frac{i}{N_\bk} \sum_{\bk\bb} c_{\abs{\bb}} \bb \, e^{-i(\bk+\bb)\cdot\bR} \, M_{ij}^{(\bk, \bb)}
    \neq \br_\ijR^{\rmSFD} \,.
\end{align}
In the third equality, we changed variables $\bb$ to $-\bb$, and then $\bk$ to $\bk + \bb$.
In the last line, the phase factor $e^{-i\bb\cdot\bR}$ that is absent in \Eq{eq:r_fd} appears, breaking Hermiticity.
For this reason, it was often necessary to perform post hoc symmetrization of the position matrix elements~\cite{2008MostofiWannier90, 2020PizziWannier90, tsirkin2021WB}.
However, when using the TEFD formula, we do not need this additional step.

We also note that the TEFD method yields size-consistent results.
See \Sec{sec:size_consistency} for a proof.


\subsection{Results} \label{sec:trinv_results}

\begin{figure}[tb]
\centering
\includegraphics[width=0.99\linewidth]{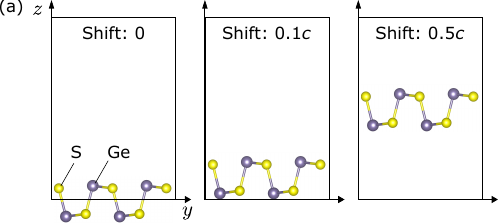}
\includegraphics[width=0.99\linewidth]{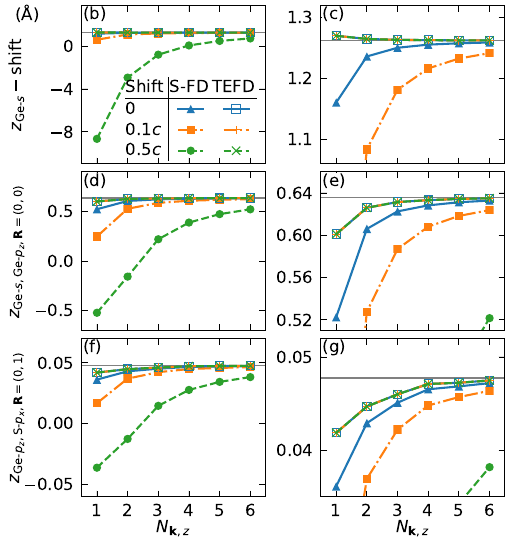}
\caption{
    (a) Three structures of monolayer GeS with different out-of-plane shifts.
    We set the out-of-plane lattice constant to $c=15~\mathrm{\AA}$.
    (b) The $z$-component of the Wannier center of the $s$-orbital-like WF of monolayer GeS for different shifts and a $12\times 12 \times N_{\bk,z}$ $\bk$ grid, computed using S-FD [\Eq{eq:center_FD}] and TEFD [\Eq{eq:tr_r_fd_self_consistent}].
    The gray horizontal line indicates the converged value at $N_{\bk,z} \to \infty$, obtained by a $1/N_{\bk,z}^2$ extrapolation.
    (c) Magnified view of panel (b).
    (d-g) Same as panels (b) and (c), but for off-diagonal matrix elements.
}
\label{fig:GeS_position}
\end{figure}

\begin{figure}[tb]
\centering
\includegraphics[width=0.99\linewidth]{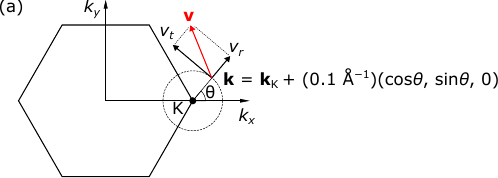}
\includegraphics[width=0.99\linewidth]{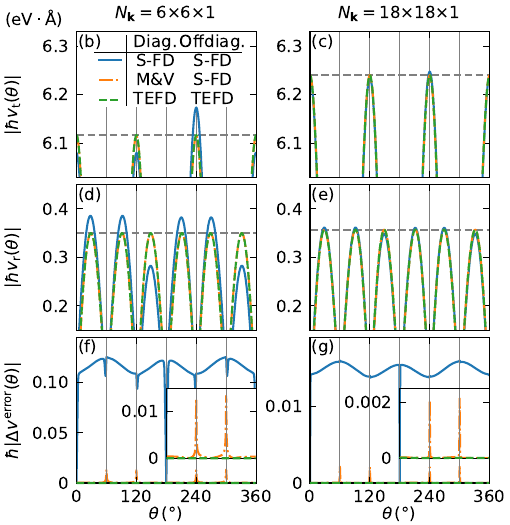}
\caption{
    (a) Brillouin zone of graphene and a circular path (dotted black circle) around the K point.
    (b, c) Tangential and (d, e) radial components of the interband velocity matrix elements between the lower and upper branches of the Dirac cone along this path, computed using \Eq{eq:velocity}.
    We show three curves, corresponding to different FD approaches for diagonal and off-diagonal position matrix elements.
    (f, g) The $C_3$ symmetry-breaking error of the interband velocity matrix elements.
    The plotted quantity is the magnitude of the difference between the velocity matrix elements at $\theta$ and $\theta + 120^\circ$ in the respective local coordinates: $\Delta v^{\rm error}(\theta) = \sqrt{\abs{v_t(\theta + 120^\circ) - v_t(\theta)}^2 + \abs{v_r(\theta + 120^\circ) - v_r(\theta)}^2}$.
    The insets in (f, g) are magnified views of the small yet finite values of the dash-dotted orange curve.
}
\label{fig:graphene_velocity}
\end{figure}

We now apply the TEFD formula [\Eq{eq:tr_r_fd_trinv}] to calculate the position matrix elements.
To demonstrate its equivariance, we consider a two-dimensional monolayer GeS (see Sec.~\ref{app:computational} for the computational details) and compute the Wannier centers and position matrix elements for three different out-of-plane shifts, as shown in \Fig{fig:GeS_position}(a).
Since all three structures are equivalent, the Wannier centers should be shifted by the same amount as the applied shift, and the off-diagonal position matrix elements should remain invariant [\Eq{eq:tr_r_transform}].
\Figus{fig:GeS_position}(b-g) show that this expectation is satisfied by the TEFD formula, while the S-FD formula [\Eq{eq:r_fd}] produces different results depending on the shift.
This result numerically demonstrates the non-equivariance of the S-FD formula and the equivariance of the TEFD formula.
The error of the S-FD formula considerably increases with the out-of-plane shift, as the WFs move away from the origin, where the S-FD works well.
We note that the M\&V [\Eq{eq:center_MV}] and the S\&S [\Eq{eq:center_SS}] centers agree with the TEFD ones, with differences smaller than $2\times10^{-4}~\mathrm{\AA}$.

In addition, when using the S-FD formula, the results converge slowly with the sampling of the $\bk$ points along the out-of-plane direction.
In calculations of a monolayer with a large vacuum, it is common to use a single $\bk$ point along the out-of-plane direction, i.e., $N_{\bk,z} = 1$.
However, for the S-FD formula, the result at $N_{\bk,z} = 1$ is far from convergence, and the error relative to the converged result at $N_{\bk,z} \to \infty$ is 8\% for the Wannier center, and 18\% and 24\% for the off-diagonal element $z_{ij\bR}$  with $ij\bR = \mathrm{Ge}\text{-}s, \mathrm{Ge}\text{-}p_z, (0, 0)$ and $\mathrm{Ge}\text{-}p_z, \mathrm{S}\text{-}p_x, (0, 1)$, respectively.
In contrast, the TEFD formula with $N_{\bk,z} = 1$ yields much smaller errors of 0.6\%, 6\%, and 12\%.
Note that the reason why one needs finite $k$-point sampling along the non-periodic $z$ direction is that the vacuum thickness is not infinitely large, and hence the system is not perfectly two-dimensional.
As the vacuum thickness in the out-of-plane direction increases, this error of the TEFD method would decrease to zero, while the error of the S-FD method would remain large depending on the position of the layer relative to the origin.

The TEFD formula is also useful for ensuring that the interpolated matrix elements respect the crystal symmetries of the system.
The S-FD method can lead to significant violations of the symmetries due to its non-equivariance.
In contrast, the TEFD formula is robust against such issues.
In Sec.~\ref{sec:symmetry}, we prove that the TEFD position matrix satisfies all crystal symmetries exactly, even in the presence of FD error, while the S-FD position matrix does not.

As an example, we compute the interband velocity matrix element, which can be expressed in terms of the position matrix elements as~\cite{2006WangAHC}
\begin{equation} \label{eq:velocity}
    \mb{v}_{mn\bk} = \frac{1}{\hbar} \nabla_\bk \bar{H}_{mn\bk} - \frac{i}{\hbar} (\veps_\nk - \veps_\mk) \mb{r}_{mn\bk} \,.
\end{equation}
Here, $\bar{H}_{mn\bk}$ is the Wannierized Hamiltonian matrix element (see Ref.~\cite{2006WangAHC} for details).

\Figu{fig:graphene_velocity} shows the interband velocity matrix elements of graphene between the lower and upper branches of the Dirac cone along a circular path centered at the K point.
These matrix elements should satisfy the $C_3$ rotational symmetry.
We compare three methods for calculating $\mb{r}_{mn\bk}$:
(i) the S-FD formula,
(ii) the M\&V formula for the diagonal part and the S-FD formula for the off-diagonals, and
(iii) the TEFD formula for the full matrix.
We find that the S-FD formula  (used by default \textsc{Wannier90} v3.1) significantly violates the $C_3$ symmetry, with an error of around 2\% for $N_\bk = 6^2$, and 0.2\% for $N_\bk = 18^2$.
Using these matrix elements to calculate physical quantities could lead to a significant violation of symmetry.
Using the equivariant M\&V formula for the Wannier centers (used when \texttt{transl\_inv = .true.} is set in \textsc{Wannier90} v3.1) reduces the error by about one order of magnitude, but the symmetry is still slightly violated.
Only when the equivariance is imposed on the full position matrix using the TEFD formula do the interband velocity matrix elements satisfy the $C_3$ symmetry.

We note that the WFs generated by the maximal localization of the total spread~\cite{1997Marzari} often may not respect the full symmetry of the crystal structure.
If this happens, the interpolated band structure and position matrix elements would not respect the symmetry.
In our calculations, we used disentanglement-only WFs without maximal localization to ensure the symmetry of the WFs and the interpolated quantities using TEFD.
One may also use symmetry-adapted WFs~\cite{2013SakumaSymmetryadapted, 2023Koretsune, 2025Oiwa} together with the TEFD method to ensure symmetric interpolation.

\section{Matrix elements of composite operators} \label{sec:composite}

\subsection{Translationally equivariant finite difference for composite operators}

Now, we generalize the TEFD method developed in the previous section to composite operators, i.e., operators that involve position operators along with other operators such as the Hamiltonian $\hat{H}$ or the spin operator $\hat{s}$.
Such composite operators are essential for computing various equilibrium and response properties.
For instance, in the case of orbital magnetization, one needs the Hamiltonian matrix elements with the $\bk$ derivative of the wavefunctions, as given by~\cite{2012LopezOrbmag}
\begin{equation} \label{eq:mathbb_B_def}
    i \mel{u^\rmW_{i\bk}}{\hat{H}_\bk}{\partial^\alpha u^\rmW_{j\bk}}
    \ \mathrm{and} \ %
    \mel{\partial^\alpha u^\rmW_{i\bk}}{\hat{H}_\bk}{\partial^\beta u^\rmW_{j\bk}} \,,
\end{equation}
where $\alpha$ and $\beta$ denote Cartesian components.
%
Similarly, for spin Hall conductivity, one needs the following matrix elements~\cite{2019RyooSHC}:
\begin{equation} \label{eq:mathbb_S_def}
    i \mel{u^\rmW_{i\bk}}{\hat{s}^\gamma}{\partial^\alpha u^\rmW_{j\bk}}
    \ \mathrm{and} \ %
    i \mel{u^\rmW_{i\bk}}{\hat{s}^\gamma \hat{H}_\bk}{\partial^\alpha u^\rmW_{j\bk}} \,,
\end{equation}
where $\hat{s}^\gamma$ is the $\gamma$-th component of the spin operator.
These quantities can be computed by Fourier transforming the real-space matrices of the form
\begin{subequations} \label{eq:real_space_def_BC}
\begin{align}
    \label{eq:real_space_def_B}
    B_{\ijR}^\alpha[\hat{O}]
    &= \mel{w_{i\mb{0}}}{\hat{O}(\hat{r} - R)^\alpha}{w_{j\bR}} \,,
    \\
    \label{eq:real_space_def_C}
    C_{\ijR}^{\alpha\beta}[\hat{O}]
    &= \mel{w_{i\mb{0}}}{\hat{r}^\alpha\hat{O}(\hat{r} - R)^\beta}{w_{j\bR}} \,,
\end{align}
\end{subequations}
where $\hat{O}$ is a general lattice-periodic operator, which does not have to be hermitian.
For orbital magnetization, we need $B[\hat{H}]$ and $C[\hat{H}]$, while for spin Hall conductivity, we need $B[\hat{s}^\gamma]$ and $B[\hat{s}^\gamma \hat{H}]$.

If the entire system is translated by $\mb{d}$, the matrix elements $B[\hat{O}]$ and $C[\hat{O}]$ transform as
\begin{subequations} \label{eq:transf_BC}
\begin{align} \label{eq:transf_B}
    B_{\ijR}^{\prime\alpha}[\hat{O}]
    &= \mel{w_{i\mb{0}}}{\hat{O}(\hat{r} + d - R)^\alpha}{w_{j\bR}}
    \nnnl
    &= B_{\ijR}^{\alpha}[\hat{O}] + d^\alpha O_{\ijR}
    \,,
    \\
    \label{eq:transf_C}
    C_{\ijR}^{\prime\alpha\beta}[\hat{O}]
    &= \mel{w_{i\mb{0}}}{(\hat{r} + d)^\alpha \hat{O}(\hat{r} + d - R)^\beta}{w_{j\bR}}
    \nnnl
    &= C_{\ijR}^{\alpha\beta}[\hat{O}] + d^\alpha B_{\ijR}^{\beta}[\hat{O}] + d^\beta B_{ji\bmR}^{\alpha*}[\hat{O}^\dagger]
    \nnnl
    &\quad + d^\alpha d^\beta O_{ij\bR}
    \,.
\end{align}
\end{subequations}
%
In the second equality of \Eq{eq:transf_C}, we used
\begin{align} \label{eq:composite_conjugate}
    \mel{w_{i\mb{0}}}{\hat{r}^\alpha \hat{O} d^\beta}{w_{j\bR}}
    &= d^\beta \mel{w_{j\bR}}{\hat{O}^\dagger \hat{r}^\alpha}{w_{i\mb{0}}}^*
    \nnnl
    &= d^\beta \mel{w_{j\mb{0}}}{\hat{O}^\dagger (\hat{r} + R)^\alpha}{w_{i\bmR}}^*
    \nnnl
    &= d^\beta B_{ji\bmR}^{\alpha*}[\hat{O}^\dagger]
    \,.
\end{align}
\Equs{eq:transf_B} and \eqref{eq:transf_C} define the equivariance condition for the composite operators.

Applying the S-FD approximation to the position operator [\Eq{eq:opr_fd}], we derive the FD expressions of $\mb{B}_{\ijR}$ and $\mb{C}_{\ijR}$~\cite{2012LopezOrbmag}:
\begin{subequations}
\begin{align} \label{eq:BB_fd}
    B_{\ijR}^{\alpha (\rmSFD)} [\hat{O}] &=
    i\sum_{\bb} c_{\abs{\bb}} b^\alpha \mel{w_{i\mb{0}}}{\hat{O} e^{-i\bb \cdot (\opbr - \bR)}}{w_{j\bR}}
    \nnnl
    &= \frac{i}{N_\bk} \sum_{\bb \bk} c_{\abs{\bb}} b^\alpha \, e^{-i\bk\cdot\bR} \, B_{ij}^{(\mb{k,b})}[\hat{O}] \,,
\end{align}
and
\begin{align} \label{eq:CC_fd}
    &\quad C_{\ijR}^{\alpha\beta (\rmSFD)} [\hat{O}] \nnnl
    &= \sum_{\bb_1 \bb_2} c_{\abs{\bb_1}} c_{\abs{\bb_2}} b_1^\alpha b_2^\beta \,
    \mel{w_{i\mb{0}}}{e^{i\bb_1 \cdot \opbr}\hat{O} e^{-i\bb_2 \cdot (\opbr - \bR)}}{w_{j\bR}}
    \nnnl &=
    \frac{1}{N_\bk} \sum_{\bb_1 \bb_2 \bk} c_{\abs{\bb_1}} c_{\abs{\bb_2}} b_1^\alpha b_2^\beta \,
    e^{-i\bk\cdot\bR} \, C_{ij}^{(\bk, \bb_1, \bb_2)}[\hat{O}]\,.
\end{align}
\end{subequations}
Here, we defined
\begin{subequations}
\begin{align}
    \label{eq:def_B_kb}
    B_{ij}^{(\bk,\bb)}[\hat{O}]
    &= \mel{u^\rmW_\ik}{\hat{O}_{\bk}}{u^\rmW_\jkb} \,,
    \\
    \label{eq:def_C_kb1b2}
    C_{ij}^{(\bk,\bb_1,\bb_2)}[\hat{O}]
    &= \mel{u^\rmW_{i\bkb_1}}{\hat{O}_{\bk}}{u^\rmW_{j\bkb_2}} \,,
\end{align}
\end{subequations}
where $\hat{O}_\bk = e^{-i \bk \cdot \opbr} \hat{O} e^{i \bk \cdot \opbr}$ acts on the periodic part of the Bloch wavefunction.

To demonstrate that these approximations are not equivariant, we analyze the leading FD error.
For $B[\hat{O}]$, we find
\begin{align} \label{eq:BB_fd_exapnd}
    &\quad B_{\ijR}^{\alpha (\rmSFD)} [\hat{O}]
    \nnnl
    &= i\sum_{\bb} c_{\abs{\bb}} b^\alpha \Bigl[-i  \expval{\hat{O} \bb \cdot (\opbr - \bR)}_{\ijR} \nnnl
    & \qquad +
    \frac{(-i)^3}{6} \expval{\hat{O} (\bb \cdot (\opbr - \bR))^3}_{\ijR} + O(b^5) \Bigr]
    \nnnl
    &= B_{\ijR}^{\alpha} [\hat{O}]
    \nnnl
    &\qquad -\frac{1}{6} \sum_{\bb} c_{\abs{\bb}} b^\alpha \expval{\hat{O} (\bb \cdot (\opbr - \bR))^3}_{\ijR}
    + O(b^4) \,.
\end{align}
Under translation, the FD error term transforms as
\begin{align}
    \expval{\hat{O} (\bb \cdot (\opbr - \bR))^3}_{\ijR}'
    &= \expval{\hat{O} (\bb \cdot (\opbr + \bd - \bR))^3}_{\ijR}
    \nnnl
    &\neq \expval{\hat{O} (\bb \cdot (\opbr - \bR))^3}_{\ijR} \,,
\end{align}
and thus \Eq{eq:transf_B} does not hold, i.\,e.\,, \Eq{eq:BB_fd} is not translationally equivariant.
Similarly, for $C[\hat{O}]$, we find
\begin{align} \label{eq:CC_fd_exapnd}
    &\quad C_{\ijR}^{\alpha\beta (\rmSFD)} [\hat{O}]
    \nnnl
    &= \sum_{\bb_1 \bb_2} c_{\abs{\bb_1}} c_{\abs{\bb_2}} b_1^\alpha b_2^\beta \Bigl[
        \expval{(\bb_1 \cdot \opbr)\hat{O} (\bb_2 \cdot (\opbr - \bR))}_{\ijR}
    \nnnl & \quad\quad
        - \frac{1}{6} \expval{(\bb_1 \cdot \opbr)^3\hat{O} (\bb_2 \cdot (\opbr - \bR))}_{\ijR}
    \nnnl & \quad\quad
        - \frac{1}{6} \expval{(\bb_1 \cdot \opbr)\hat{O} (\bb_2 \cdot (\opbr - \bR))^3}_{\ijR}
        + O(b^6)
    \Bigr]
    \nnnl
    &= C_{\ijR}^{\alpha\beta} [\hat{O}]
    \nnnl
    &\qquad - \frac{1}{6} \sum_{\bb_1} c_{\abs{\bb_1}} b_1^\alpha \expval{(\bb_1 \cdot \opbr)^3 \hat{O} (\hat{r} - R)^\beta}_{\ijR}
    \nnnl
    &\qquad - \frac{1}{6} \sum_{\bb_2} c_{\abs{\bb_2}} b_2^\beta \expval{\hat{r}^\alpha \hat{O} (\bb_2 \cdot (\opbr - \bR))^3}_{\ijR}
    \nnnl
    &\qquad + O(b^4)
    \,.
\end{align}
Once again, the FD error is not translationally equivariant.

Following the same procedure for position matrix elements, we now derive TEFD expressions for the composite operators.
We first add constant shifts to $\opbr$ in \Eq{eq:BB_fd} and \Eq{eq:CC_fd} and define the ``centered'' matrix elements as
\begin{subequations}
\begin{align}
    \label{eq:def_BB_tilde}
    \tilde{B}_{\ijR}^{\alpha}[\hat{O}]
    &= \mel{w_{i\mb{0}}}{\hat{O}(\hat{r}-\bar{r}_{\ijR})^\alpha}{w_{j\bR}} \,
    \\
    \label{eq:def_CC_tilde}
    \tilde{C}_{\ijR}^{\alpha\beta}[\hat{O}]
    &= \mel{w_{i\mb{0}}}{(\hat{r}-\bar{r}_{\ijR})^\alpha \hat{O}(\hat{r}-\bar{r}_{\ijR})^\beta}{w_{j\bR}} \,.
\end{align}
\end{subequations}
Here, $\bar{r}_{\ijR}$ is the midpoint of the two Wannier centers, as defined in \Eq{eq:tr_r_midpoint_def}.
Next, we apply the TEFD representation of the position operator, \Eq{eq:opr_TEFD}, and derive the TEFD expressions:
\begin{subequations}
\begin{align}
    \label{eq:BB_tilde_FD}
    \tilde{B}_{\ijR}^{\alpha (\rmTEFD)} [\hat{O}]
    &= i \sum_{\bb} c_{\abs{\bb}} b^\alpha
    \\
    &\ \times \mel{w_{i\mb{0}}}{\hat{O} e^{-i\bb \cdot (\opbr - \bar{\br}_{\ijR})}}{w_{j\bR}}
    \,,
    \nnnl
    \label{eq:CC_tilde_FD}
    \tilde{C}_{\ijR}^{\alpha\beta (\rmTEFD)}[\hat{O}]
    &= \sum_{\bb_1 \bb_2} c_{\abs{\bb_1}} c_{\abs{\bb_2}} b_1^\alpha b_2^\beta
    \\
    &\ \times \mel{w_{i\mb{0}}}{e^{i\bb_1 \cdot (\opbr - \bar{\br}_{\ijR})}\hat{O}e^{-i\bb_2 \cdot (\opbr - \bar{\br}_{\ijR})}}{w_{j\bR}}
    \nonumber
    \,.
\end{align}
\end{subequations}
Finally, using the Bloch functions basis, we obtain:
\begin{subequations}
\begin{align} \label{eq:BB_TI_FD}
    & \quad \tilde{B}_{\ijR}^{\alpha (\rmTEFD)} [\hat{O}]
    \nnnl
    &= \frac{i}{N_\bk} \sum_{\bb \bk} c_{\abs{\bb}} b_{\alpha} \,
    e^{i\bb \cdot \bar{\br}_{\ijR}} e^{- i (\bk + \bb) \cdot \bR}
    \mel{\psi^\rmW_\ik}{\hat{O}e^{-i\bb\cdot\opbr}}{\psi^\rmW_\jkb}
    \nnnl
    &= \frac{i}{N_\bk} \sum_{\bb \bk} c_{\abs{\bb}} b^\alpha \,  e^{i\bb \cdot (\bar{\br}_{\ijR} - \bR)}
    e^{-i \bk \cdot \bR}
    B^{(\bk,\bb)}_{ij}[\hat{O}]
\end{align}
and
\begin{align} \label{eq:CC_TI_FD}
    & \quad \tilde{C}_{\ijR}^{\alpha\beta (\rmTEFD)} [\hat{O}]
    \nnnl
    &= \frac{1}{N_\bk} \sum_{\bb_1 \bb_2 \bk}  c_{\abs{\bb_1}} c_{\abs{\bb_2}} b_1^\alpha b_2^\beta \,
    e^{-i\bb_1 \cdot \bar{\br}_{\ijR}} e^{i\bb_2 \cdot \bar{\br}_{\ijR}} e^{- i (\bk + \bb_2) \cdot \bR}
    \nnnl
    &\qquad \times
    \mel{\psi^\rmW_{i\bk + \bb_1}}{e^{i\bb_1\cdot\opbr}\hat{O}e^{-i\bb_2\cdot\opbr}}{\psi^\rmW_{j\bk+\bb_2}}
    \nnnl
    &= \frac{1}{N_\bk} \sum_{\bb_1 \bb_2 \bk} c_{\abs{\bb_1}} c_{\abs{\bb_2}} b_1^\alpha b_2^\beta \, e^{-i\bb_1 \cdot \bar{\br}_{\ijR}} e^{i\bb_2 \cdot (\bar{\br}_{\ijR} - \bR)}
    e^{-i \bk \cdot \bR}
    \nnnl
    &\qquad \times
    C^{(\bk,\bb_1,\bb_2)}_{ij}[\hat{O}] \,.
\end{align}
\end{subequations}

We can obtain the original matrix elements in \Eqs{eq:real_space_def_B} and \eqref{eq:real_space_def_C} from \Eqs{eq:def_BB_tilde} and \eqref{eq:def_CC_tilde}:
\begin{subequations}
\begin{equation} \label{eq:B_recenter}
    B_{\ijR}^{\alpha} [\hat{O}] = \tilde{B}_{\ijR}^{\alpha} [\hat{O}] + (\bar{r}_{\ijR} - R)^\alpha O_{\ijR} \,,
\end{equation}
and
\begin{align} \label{eq:C_recenter}
    C_{\ijR}^{\alpha \beta}
    &=
    \tilde{C}_{\ijR}^{\alpha \beta}
    + \bar{r}_{\ijR}^\alpha \tilde{B}_{\ijR}^{\beta} [\hat{O}]
    + B^{\alpha *}_{ji\bmR}[\hat{O}^\dagger] (\bar{r}_{\ijR} - R)^\beta
    \nnnl
    &= \tilde{C}_{\ijR}^{\alpha \beta}
    + \bar{r}_{\ijR}^\alpha B_{\ijR}^{\beta} [\hat{O}]
    -\bar{r}_{\ijR}^\alpha (\bar{r}_{\ijR} - R)^\beta O_{\ijR}
    \nnnl
    &\quad  + {B}_{ji\bmR}^{\alpha*} [\hat{O}^\dagger]
    (\bar{r}_{\ijR} - R)^\beta \,.
\end{align}
\end{subequations}
We used \Eq{eq:composite_conjugate} to obtain the last term in the first equality of \Eq{eq:C_recenter}.
By rearranging the terms and returning to the original matrix elements, we arrive at the TEFD expressions for the composite matrix elements:
\begin{widetext}
\begin{subequations}
\begin{align} \label{eq:BB_trinv}
    B_{\ijR}^{\alpha (\rmTEFD)} [\hat{O}]
    &= \frac{i}{N_\bk} \sum_{\bb\bk} c_{\abs{\bb}} b^\alpha \,
    e^{i\bb \cdot (\bar{\br}_{\ijR}-\bR)}
    e^{-i\bk\cdot\bR} B_{ij}^{(\mb{k,b})}[\hat{O}] + (\bar{r}_{\ijR} - R)^\alpha O_{\ijR}
    \\
    \label{eq:CC_trinv}
    C_{\ijR}^{\alpha\beta (\rmTEFD)} [\hat{O}]
    &= \frac{1}{N_\bk} \sum_{\bb_1 \bb_2 \bk} c_{\abs{\bb_1}} c_{\abs{\bb_2}} b_{1}^\alpha b_{2}^\beta \,
    e^{-i\bb_1 \cdot \bar{\br}_{\ijR}} e^{i\bb_2 \cdot (\bar{\br}_{\ijR}-\bR)}
    e^{-i\bk\cdot \bR}
    C_{ij}^{(\bk,\bb_1,\bb_2)}[\hat{O}]
    \nnnl
    &\quad + \bar{r}_{\ijR}^\alpha B_{\ijR}^{\beta (\rmTEFD)} [\hat{O}]
    + B_{ji\bmR}^{\alpha* (\rmTEFD)} [\hat{O}^\dagger](\bar{r}_{\ijR} - R)^\beta
    - \bar{r}_{\ijR}^\alpha (\bar{r}_{\ijR} - R)^\beta O_{ij\bR} \,.
\end{align}
\end{subequations}
\end{widetext}
For other types of composite operators, one can follow the same procedure to derive the corresponding TEFD formula.

\begin{figure}[tb]
\centering
\includegraphics[width=0.99\linewidth]{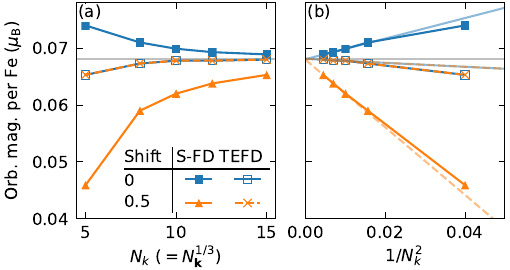}
\caption{
    Orbital magnetization of bcc Fe, computed using the S-FD and TEFD methods, as a function of (a) the coarse grid size and (b) its squared inverse.
    The results for two equivalent but shifted structures are shown.
    The shift is applied along all three directions: for a shift of $0.5$, the Fe atom is centered at $(0.5, 0.5, 0.5)$ in crystal coordinates.
}
\label{fig:orb_mag_Fe}
\end{figure}

\subsection{Orbital magnetization and spin Hall conductivity} \label{eq:composite_results}

\begin{figure}[tb]
\centering
\includegraphics[width=0.99\linewidth]{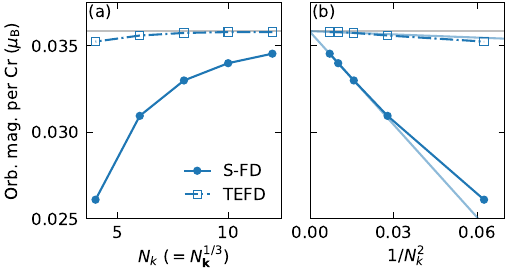}
\caption{
    Orbital magnetization of CrO$_2$, computed using the S-FD and TEFD methods, as a function of (a) the coarse grid size and (b) its squared inverse. The Cr atoms are at (0.0, 0.0, 0.0) and (0.5, 0.5, 0.5) in crystal coordinates.
}
\label{fig:orb_mag_CrO2}
\end{figure}

\begin{figure}[tb]
\centering
\includegraphics[width=0.99\linewidth]{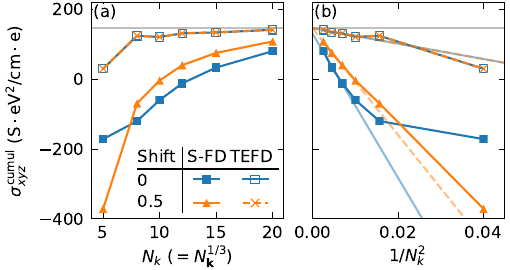}
\caption{
    Cumulative spin Hall conductivity [\Eq{eq:SHC_cumulative}] of GaAs, computed using the S-FD and TEFD methods, as a function of (a) the coarse grid size and (b) its squared inverse.
    The results for two equivalent but shifted structures are shown.
    Before the shift, the Ga and As atoms are at $(0.0, 0.0, 0.0)$ and $(-0.25, -0.25, -0.25)$, respectively, in crystal coordinates.
    After the shift of $0.5$, the two atoms are placed at $(0.5, 0.5, 0.5)$ and $(0.25, 0.25, 0.25)$.
}
\label{fig:SHC}
\end{figure}

We now apply the TEFD method to compute the orbital magnetization of bcc Fe and CrO$_2$ and the spin Hall conductivity of GaAs.
First, for bcc Fe, we consider two equivalent but translated crystal structures.
The first one has an Fe atom at the origin, while the second one has an Fe atom at $(0.5,\, 0.5,\, 0.5)$ in crystal coordinates.
Since the two structures are equivalent, they should have the same orbital magnetization.
However, as shown in \Fig{fig:orb_mag_Fe}, this is only the case for the TEFD method, not the S-FD method.
Moreover, while the two methods converge to the same value as the coarse $\bk$-grid density increases, the TEFD method converges much faster.
%
%
Although placing the Fe atom at the origin reduces the error for the S-FD method, the TEFD method still outperforms it because real-space off-diagonal matrix elements are treated differently.
This result demonstrates that simply replacing the S-FD method with the TEFD method can significantly improve the precision of orbital magnetization calculations.

We obtain similar results for the orbital magnetization of CrO$_2$.
As shown in \Fig{fig:orb_mag_CrO2}, the TEFD method converges much faster than the S-FD method.
Since CrO$_2$ has multiple atoms in the unit cell, there are always atoms that are not at the origin of the unit cell.
As a result, the large error of the S-FD method is unavoidable, whereas the use of the TEFD method provides a significant improvement.


Finally, we compute the spin Hall conductivity of GaAs using the TEFD method.
We used the method of \citet{2019RyooSHC} to compute the spin Hall conductivity.
(See Sec.~\ref{sec:spin_vel} for how we avoided an unnecessary approximation of this method. We find that the effect of the approximation is small in the case of GaAs.)
To quantify the convergence of the frequency-dependent spin Hall conductivity with a single number, we define the cumulative spin Hall conductivity:
\begin{equation} \label{eq:SHC_cumulative}
    \sigma_{xyz}^{\text{cumul}} = \nbint{0}{\omega_{\max}} \dd \omega \, \hbar \omega \Im \sigma_{xyz}(\omega) \,,
\end{equation}
where $\sigma_{xyz}(\omega)$ is the frequency-dependent spin Hall conductivity.
We set $\hbar \omega_{\rm max} = 20~\mathrm{eV}$.
We numerically evaluate \Eq{eq:SHC_cumulative} and analyze its convergence with respect to the coarse grid size.
We note that in the limit $\omega_{\rm max} \to \infty$, the cumulative spin Hall conductivity can be written in terms of matrix elements between only the occupied states, without the need for a frequency integral (see \Sec{app:cumulative_shc}).

The convergence of the cumulative spin Hall conductivity of GaAs is shown in \Fig{fig:SHC}.
As in the orbital magnetization examples, the TEFD method significantly outperforms the S-FD method in computing spin Hall conductivity, both in the invariance with respect to translations and in the magnitude of the FD error.

\section{Higher-order finite difference} \label{sec:higher_order}

We now turn to the second key result of this paper: the higher-order FD (HOFD) method for position and composite operators.
Thanks to the localization of the WFs, the interpolation error for energy eigenvalues decays nearly exponentially with increasing $\bk$-grid density.
However, if we use the FD (even TEFD) method, the errors for position and composite operators decay only quadratically with respect to $b$, the distance between the neighboring points in the coarse \bk-point grid.
HOFD provides a way to accelerate this error scaling.
Although the use of HOFD for position matrix elements has been suggested previously~\cite{1997Marzari, 2008MostofiWannier90, 2012LopezOrbmag}, it has not been systematically implemented.
Here, we present the HOFD method in detail, focusing in particular on the construction of the $\bb$ vectors and their associated weights.
We demonstrate both formally and numerically that the HOFD method significantly improves convergence, changing the scaling exponent for $b$ of the FD error from $-2$ to $-2N$, where $N$ ($>1$) is the order of HOFD.

\subsection{\texorpdfstring{$\bb$}{b} vectors and weights for higher-order finite difference} \label{sec:hofd_bvectors}

\begin{table}[]
\begin{tabular}{c||c|cc|cc}
Order               & \multirow{2}{*}{1}  & \multicolumn{2}{c}{2} & \multicolumn{2}{c}{3} \\ \cline{1-1} \cline{3-6}
Method              &    & Shells  & Multiples   & Shells  & Multiples  \\ \hline\hline
Simple cubic        & 6  & 24      & 12          & 86      & 18         \\
Face-centered cubic & 8  & 26      & 16          & 82      & 24         \\
Body-centered cubic & 12 & 42      & 24          & 86      & 36         \\
Hexagonal           & 8  & 28      & 16          & 72      & 24         \\
Tetragonal          & 6  & 24      & 12          & 62      & 18
\end{tabular}
\caption{
    The number of $\bb$ vectors used for the multiples and shells schemes for HOFD with different Bravais lattices.
}
\label{tab:bvectors}
\end{table}

The first-order FD scheme is governed by the conditions in \Eq{eq:c_b_sum} for the $\bb$ vectors and their associated weights.
A natural extension is to enforce additional constraints with a sufficiently large collection of $\bb$ vectors and weights.
For the $N$th-order FD scheme, we arrive at the following conditions, the first line of which is \Eq{eq:c_b_sum}:
\begin{align}
\label{eq:fd_nearest_nd}
\begin{split}
    &\sum_{\bb}c_{\abs{\bb}}b^\ai b^\aii=\delta_{\ai \aii}\\
    &\sum_{\bb} c_{\abs{\bb}} b^\ai b^\aii b^{\alpha_3} b^{\alpha_4} = 0 \\
    &\,\,\,\,\,\vdots\\
    &\sum_{\bb} c_{\abs{\bb}} b^\ai b^\aii \cdots b^{\alpha_{2N}} = 0 \,.
\end{split}
\end{align}
A full derivation is provided in Sec.~\ref{app:derivation_hofd}.
If these conditions are satisfied, the FD error scales as $O(b^{2N})$, see \Eq{eq:hofd_error}.

The set of $\bb$ vectors and weights $c_{\abs{\bb}}$ that satisfy \Eq{eq:fd_nearest_nd} can be found using the same algorithm as the first-order case~\cite{2008MostofiWannier90}.
One first groups all possible $\bb$ vectors into shells of equal distances from the origin.
Then, starting with the shell of the smallest radius, one checks if the linear equation in \Eq{eq:fd_nearest_nd} for the weights $c_{\abs{\bb}}$ has a solution.
If no solution is found, we iteratively add shells, starting from smaller radii and increasing outward, until the conditions, \Eq{eq:fd_nearest_nd} are satisfied.
This scheme yields a set of $\bb$ vectors with the smallest maximum radius.
We call this the ``shells'' method for HOFD.
This method was recently applied to the calculation of $\bk$ gradients using second-order HOFD~\cite{Cistaro2023}.

However, the shells method has a problem: the number of selected $\bb$ vectors grows very rapidly with $N$.
Each line of \Eq{eq:fd_nearest_nd} contains $6$, $15$, $28$, $45$ $\cdots$, $\multiset{3}{2N} = (2N+1)(N+1)$ non-equivalent equations.
Since there are $N$ lines, the total number of constraints grows as $6$, $6+15=21$, $21+28=49$, $49+45=94$, $\cdots$, $O(N^3)$.
Therefore, it is natural that the total number of $\bb$ vectors, which seemingly have to be larger than the total number of constraints, increases rapidly with $N$.
This is the case in the shells method: note that the number of $\bb$ vectors for $N=2$ and $N=3$ shells methods is larger than 21 and 49, respectively (Tab.~\ref{tab:bvectors}).
For the simple cubic lattice, the third-order HOFD method uses 86 $\bb$ vectors, 14.3 times as many as the 6 $\bb$ vectors of first-order FD.
Table~\ref{tab:bvectors} shows the number of $\bb$ vectors for several representative Bravais lattices.
Moreover, as the radius of the shell increases, the distance between neighboring shells is reduced, which can make it hard or complicated to numerically distinguish different shells at high $N$.

To address this problem, we present an alternative method for finding the $\bb$ vectors for HOFD: the ``multiples'' method.
The key idea is that the $N$th-order HOFD constraint [\Eq{eq:fd_nearest_nd}] can be satisfied by using the first to $N$th multiples of each of the $\bb$ vectors obtained by the usual first-order FD method:
\begin{equation} \label{eq:hofd_multiples_b}
    \bigl\{\bk+\bb,\; \bk+2\bb,\; \dots,\; \bk+N\bb\bigr\} \,.
\end{equation}
The corresponding weight for $\bk+m\bb$, $c_{\abs{m\bb}}^{(N)}$, can be determined by solving a one-dimensional HOFD problem, and the analytic solution is given by:
\begin{equation}
    c_{\abs{m\bb}}^{(N)} = c_{\abs{\bb}}^{(1)} \frac{1}{m^2} \prod_{n \neq m}^{N} \frac{n^2}{n^2 - m^2}\,.
\label{eq:cramers_nd_}
\end{equation}
Here, $c_{\abs{\bb}}^{(1)}$ is the first-order coefficient.
See \Sec{app:derivation_hofd} for the derivation of \Eq{eq:cramers_nd_}.

The $N$th-order HOFD method with the multiples method uses $N \times N_{\rm 1FD}$ vectors, where $N_{\rm 1FD}$ is the number of $\bb$ vectors in the first-order FD case.
This number scales as $O(N)$, which is much more favorable than the shells method.
Table~\ref{tab:bvectors} compares the numbers of $\bb$ vectors needed by the two approaches for second- and third-order FD methods.
The multiples method significantly reduces both computational time and storage relative to the shells method.

\begin{figure}[tb]
\centering
\includegraphics[width=0.99\linewidth]{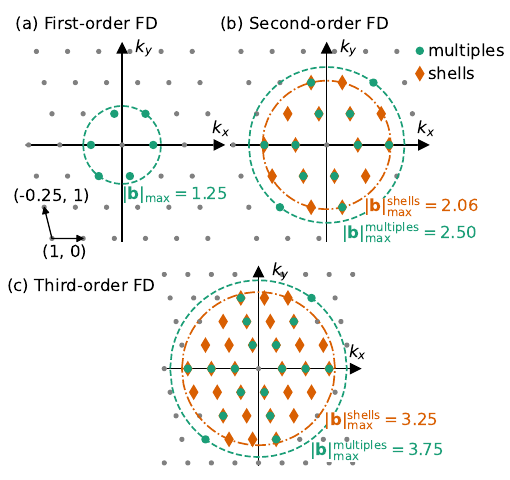}
\caption{
    (a) Selected $\bb$ vectors for the first-order FD method applied to a two-dimensional system with reciprocal lattice vectors $(1, 0)$ and $(-0.25, 1)$.
    (b) Selected $\bb$ vectors for the second-order FD method using the multiples scheme (green circles) and the shells scheme (orange diamonds).
    The $\bb$ vectors for the multiples scheme are integer multiples of the $\bb$ vectors in (a).
    The $\bb$ vectors for the shells scheme fit inside a circle of smaller radius ($2.06$) than that for the multiples scheme ($2.50$).
    We note that for a square lattice, the two schemes are equivalent.
}
\label{fig:higher_order_search}
\end{figure}

\begin{figure*}[tb]
\centering
\includegraphics[width=0.99\linewidth]{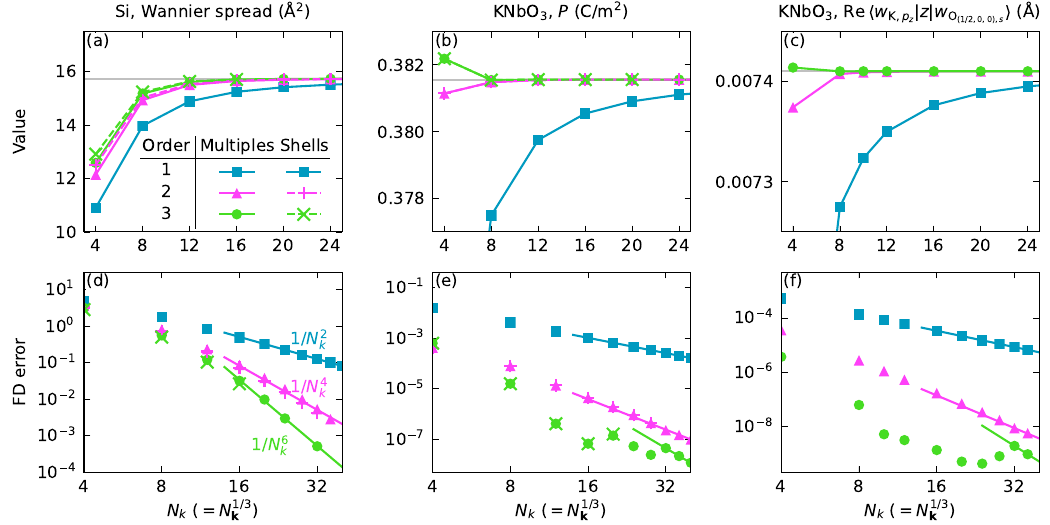}
\caption{
    Convergence of (a) the Wannier spread of Si,
    (b) the macroscopic polarization [\Eq{eq:wannier_polarization}] of KNbO$_3$, and
    (c) the off-diagonal $z$ matrix element between the WFs constructed from a $p_z$ orbital at a potassium atom and an $s$ orbital at an oxygen atom of KNbO$_3$.
    We used uniform coarse $\bk$-point grids of size $N_\bk = N_k \times N_k \times N_k$.
    For all cases, we used equivariant formulas.
    For (a, b), we compare the results from the multiples and shells schemes for HOFD and find that the two results are almost identical except at very coarse meshes.
    Thus, in (c), we show only the results from the multiples scheme.
    Note that the two schemes are equivalent for the first-order FD method.
    The converged polarization of KNbO$_3$ [panel (b)] agrees well with a previous calculation of Ref.~\cite{2006Stengel}.
    (d)--(f) FD error plotted on a log-log scale, showing the expected scaling $O(1/N_k^{2N})$, represented by lines.
}
\label{fig:higher_convergence}
\end{figure*}

Figure~\ref{fig:higher_order_search} illustrates the first- and second-order choices of $\bb$ vectors for a two-dimensional example.
The multiples method employs $\bb$ vectors with a larger average length than those used in the shells method.
Also, the vector distribution is less spherically symmetric than that of the shells method.
Consequently, the prefactor of the nominal $O(b^{2N})$ error can be larger.
It is valuable to numerically test whether this difference is significant enough to justify the use of the shells method despite the larger number of $\bb$ vectors.
In the next section, we perform this test and find that the errors of the two methods are very similar (see Fig.~\ref{fig:higher_convergence}).
%

We note that there are different options in choosing the $\bb$ vectors even for the first-order FD method~\cite{1997Marzari, Silvestrelli1999, 2008MostofiWannier90}.
Choosing shells with the smallest radii does not necessarily lead to the most spherically symmetric FD approximation~\cite{Posternak2002}.
Spherical symmetry may be important when dealing with the position matrix elements between atomic-orbital-like WFs, which exhibit approximate angular momentum selection rules.
In such cases, the first-order $\bb$ vectors can be manually selected~\cite{Posternak2002, 2008MostofiWannier90}, and then extended to higher orders using the multiples scheme.

One may wonder why the first-order FD method requires six, instead of four, $\bb$ vectors in the lattice of \Fig{fig:higher_order_search}.
In fact, using four lattice vectors inside the circle of \Fig{fig:higher_order_search}(a) is sufficient to form a first-order FD approximation.
However, such an FD scheme is \textit{not} described by the conventional parametrization \Eq{eq:fd_naive}.
One needs to use a generalized scheme, where one has a degree of freedom to choose vector-valued coefficients instead of the scalar coefficients $c_{\abs{\bb}}$.
See \Sec{sec:alternative_fd} for more details about this vector-coefficient FD scheme.

The HOFD method can be easily combined with the TEFD formalism to form the TE-HOFD method.
In fact, using TEFD formulas is essential for the effectiveness of the HOFD method.
In \Figs{fig:schematic}(c, d), we schematically illustrate the HOFD method.
As the FD order increases, the sinusoidal approximation evolves into a sawtooth-like function.
In other words, the HOFD scheme can be understood as constructing an improved approximation of the sawtooth function using a linear combination of a few sine functions, as illustrated in \Fig{fig:schematic}(c, d)~\footnote{The method of Ref.~\cite{2006Stengel}, which uses a sawtooth wave to calculate the Wannier centers, can be regarded as an infinite-order HOFD.}.
However, because the sawtooth-like approximation is still periodic, it accurately represents the aperiodic position operator well only close to the center of the sawtooth.
Although higher-order approximations improve the linearity, they inevitably break down near the periodic boundaries.
In the S-FD approach, the center of the sawtooth is fixed at the origin, which may not coincide with regions where the WFs have significant weight.
Only when combined with the TEFD formalism, which shifts the center of the approximation to better align with the relevant WFs, does HOFD enable accurate computation of the position and composite matrix elements.

The same TEFD formula for the position matrix [\Eq{eq:tr_r_fd_trinv}] and the composite operators [\Eqs{eq:BB_trinv} and \eqref{eq:CC_trinv}] can be used for TE-HOFD, just by replacing the $\bb$ vectors and the corresponding first-order FD coefficients, $c_{\abs{\bb}}$'s, with the HOFD ones.
The TE-HOFD method provides the most accurate and efficient way to Wannier interpolate the position and composite operators.

\subsection{Results}

We first test the convergence of the HOFD method and compare the shells and multiples schemes.
We calculate three quantities.
First, we calculate the Wannier spread of Si, which measures the localization of the WFs.
Second, we calculate the electric polarization of a cubic insulator, KNbO$_3$.
The electric polarization is given by the sum of the ionic and electronic contributions, where the electronic part is the sum of the Wannier centers of occupied states~\cite{1993KingSmithModernPol,1993RestaModernPol, Vanderbilt1993, 1994RestaRMP}:
\begin{equation}
    {\bf P} = \frac{e}{V} \Bigl( \sum_{I} Z_{I}{\bf r}_{I}
    - 2\sum_{i \in {\rm occ.}}{\bf r}_i \Bigr) \,.
\label{eq:wannier_polarization}
\end{equation}
Here, $I$ is an ionic index, $Z_I e$ is the charge of the $I$-th ion, and $\br_{I}$ is its position.
The factor of 2 accounts for spin degeneracy in the electronic contribution.
Finally, we calculate the off-diagonal position matrix elements of KNbO$_3$.
In all cases, TEFD formulas are used.

Figures~\ref{fig:higher_convergence}(a)--(c) compare these quantities computed using the first-, second-, and third-order FD methods.
In particular, we examine how the values converge as we increase $N_k$, the number of $\bk$ points along each direction.
We find that HOFD converges on much coarser $\bk$-point grids than the first-order FD method for all three quantities.
The improvement from the second- to third-order FD method is less significant than the improvement from the first- to second-order FD method on the linear scale.

Comparing the multiples and shells methods, we find little difference in performance, even though the latter uses significantly more $\bb$ vectors than the former.
The number of $\bb$ vectors used in the two methods is shown in Tab.~\ref{tab:bvectors}.
For the third-order HOFD calculation of KNbO$_3$, the multiples method uses 18 $\bb$ vectors, while the shells method uses 62 $\bb$ vectors, 3.44 times more than the multiples method.
Given the comparable accuracy and the much more straightforward implementation, we conclude that the multiples method is preferable to the shells method.

The linearity of the FD error shown in \Figs{fig:higher_convergence}(d)--(f) on the log-log scale confirms the $1/N_k^{2N}$ scaling of the FD error for the $N$th-order FD method (equivalently, the scaling is $O(b^{2N})$, where $b\propto 1/N_k$).
The slopes of the lines are $-2$, $-4$, and $-6$, for the first-, second-, and third-order FD methods, respectively.
We note that for the third-order FD results in \Figs{fig:higher_convergence}(f), the asymptotic scaling regime is reached only for $N_\bk \geq 28$.

The computational bottleneck in Wannier interpolation of physical quantities typically lies in either the non-self-consistent calculation of wavefunctions on a coarse $\bk$-point grid or the subsequent interpolation to a dense one.
In contrast, the evaluation of real-space matrix elements using FD or HOFD expressions is comparatively inexpensive.
Because our HOFD method can significantly reduce the number of $\bk$-points required for convergence, it provides a substantial overall speedup.
Based on our results, we recommend using the second-order FD method with $\bb$ vectors and $c_{\abs{\bb}}$'s determined by the multiples method, which offers the best balance between accuracy and simplicity.

\begin{figure}[tb]
\centering
\includegraphics[width=0.99\linewidth]{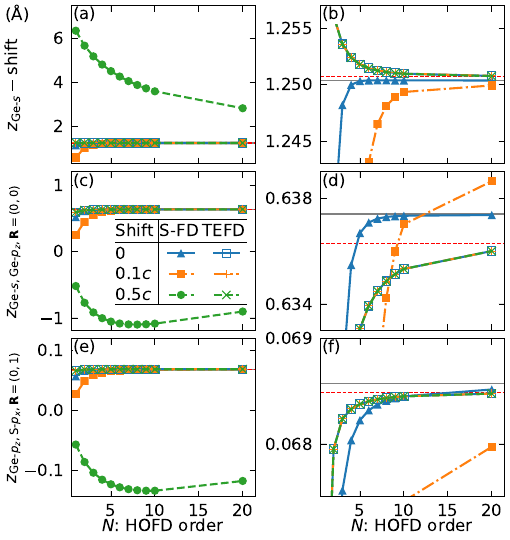}
\caption{
    Wannier centers and position matrix elements of monolayer GeS (see \Fig{fig:GeS_position} for details), computed using (i) S-FD plus HOFD method and (ii) TE-HOFD method as a function of order.
    We used a $12\times12\times1$ $\bk$ grid.
    The horizontal lines show the results obtained using a direct real-space calculation using the full FFT mesh of the plane-wave basis DFT calculation.
    The $z$ operator is approximated as a sawtooth function, with the center at the origin of the unit cell (solid gray lines) or at the Wannier center or the midpoint of the two Wannier centers (dashed red lines).
}
\label{fig:GeS_position_hofd}
\end{figure}

\begin{figure}[tb]
\centering
\includegraphics[width=0.99\linewidth]{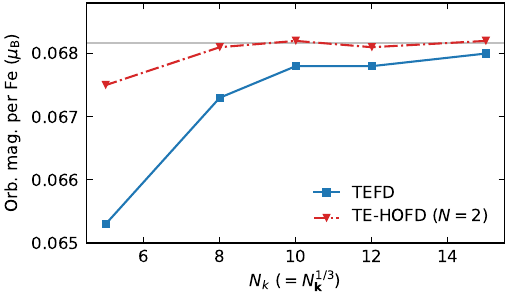}
\caption{
    Orbital magnetization of bcc Fe as a function of the coarse grid, computed using the first-order TEFD and second-order TE-HOFD methods.
}
\label{fig:orb_mag_higher}
\end{figure}

\begin{figure}[tb]
\centering
\includegraphics[width=0.99\linewidth]{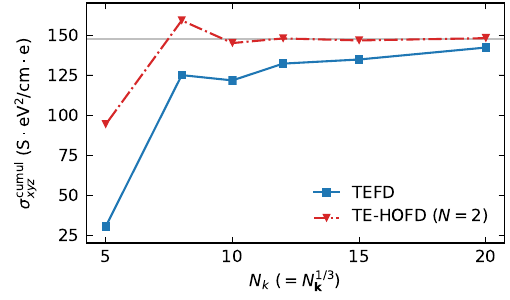}
\caption{
    Cumulative spin Hall conductivity [\Eq{eq:SHC_cumulative}] of GaAs as a function of the coarse grid, computed using the first-order TEFD and second-order TE-HOFD methods.
}
\label{fig:shc_higher}
\end{figure}

Next, we analyze how HOFD affects the matrix elements of monolayer GeS.
In the simulation of two-dimensional materials using periodic boundary conditions, a large yet finite vacuum thickness is used in the out-of-plane direction.
This leads to a slow convergence in the FD calculations of the position matrix elements in the out-of-plane direction with respect to the vacuum thickness.
One solution is to use an exact calculation using the sawtooth representation of the position operator~\cite{2006Stengel}, evaluated in the plane-wave basis.
Here, we show that this exact calculation, which amounts to an infinite-order HOFD, can be well approximated by the TE-HOFD method with a finite order.

\Figu{fig:GeS_position_hofd} shows that the TEFD method combined with the HOFD method rapidly converges to the exact result, with error below 1\% at the fourth order.
However, the S-FD converges well only if the shift is small.
We note that the matrix elements of the sawtooth function depend on its center, so the S-FD results at zero shift converge to a different value than the TEFD results.
We also note that since the vacuum thickness is finite and the electronic wavefunction have nonzero weights at the discontinuity of the sawtooth function, the result obtained in the $N \to \infty$ limit does not agree with the $N_{k,z} \to \infty$ limit of \Fig{fig:GeS_position}.
The two limits will agree only when the vacuum thickness is infinite.





Finally, we apply the HOFD method to calculate the orbital magnetization and spin Hall conductivity.
We compare the second-order HOFD method using the multiples scheme with the first-order one.
\Figu{fig:orb_mag_higher} shows that the HOFD method significantly improves the convergence of the orbital magnetization of bcc Fe with respect to the density of the $\bk$-point grid.
The error from the second-order FD method at an $8^3$ $\bk$-grid is already smaller than that from the first-order one at a $15^3$ grid.
A similar improvement is observed for the spin Hall conductivity of GaAs, as shown in Fig.~\ref{fig:shc_higher}.

\section{Vector-coefficient finite difference method} \label{sec:alternative_fd}

So far, the discussed FD methods, including S-FD, TEFD, and HOFD, were all based on the assumption in Eq.~\eqref{eq:fd_naive}~\cite{1997Marzari}.
However, this condition is not a requirement for FD methods.
In this section, we present an alternative FD scheme that does not enforce this unnecessary condition and has certain advantages.
We call it the \textit{vector-coefficient finite difference} method, as the FD coefficients are vector-valued instead of being scalar-valued.

\subsection{First-order FD and HOFD methods}

Suppose that we have a \textit{minimal number}, $2N_\bb$, of $\bb$ vectors satisfying the condition that (i) not all the $\bb$ vectors are coplanar:  $\{\bb_1,\,\, -\bb_1,\,\, \bb_2,\,\, -\bb_2,\,\, ...\,\,,\,\, \bb_{N_\bb}, -\bb_{N_\bb}\}$ to ensure the FD solution of the gradient [\Eq{eq:Bx_eq_c_sol}] and (ii) all the $\bb$ vectors of one or more shells (each shell has $\bb$ vectors of the same magnitude) are included to preserve the symmetry properties of the gradient [see Eqs.~\eqref{eq:grad_f_symm}, \eqref{eq:grad_f_symm_FD3}, and~\eqref{eq:new_method_qSb}].
Normally, we choose the shells of $\bb$ vectors with minimal lengths.
(In case the Brillouin zone is highly anisotropic, the shells can be selected differently to make the $\bb$ vector distribution more spherically symmetric.)
Then, we can approximate $\bb_i\cdot\nabla_\bk f(\bk)$ $(i=1,\,\,2,\,\,...\,\,,\,\,N_\bb)$ by
\begin{align}
\label{eq:fd_new_method}
\begin{split}
    \bb_i\cdot\nabla_\bk f(\bk)
    \approx \sum_{m=1}^N\tilde{c}_{m}^{(N)} \left[f(\bk+m\bb_i)-f(\bk-m\bb_i)\right]
\end{split}
\end{align}
in the first-order ($N=1$) or higher-order ($N>1$) FD scheme. Note that in the HOFD case, we choose $m\bb_i's$ ($m=1,\,2,\,\,...\,,N$) for the $\bb$ vectors, as in the multiples scheme with the constraint in Eq.~\eqref{eq:fd_naive} (Sec.~\ref{sec:higher_order}).
Now, we do a Taylor expansion and obtain
\begin{align}
\label{eq:fd_new_method_taylor_exp}
    &\quad \sum_{m=1}^N\tilde{c}_{m}^{(N)} \left[f(\bk+m\bb_i)-f(\bk-m\bb_i)\right]\nnnl
    &= 2\sum_{m=1}^N\tilde{c}_{m}^{(N)} \sum_{n=1}^\infty \frac{m^{2n-1}}{(2n-1)!}(\bb_i\cdot\nabla_\bk)^{2n-1}f(\bk)\nnnl
    &= 2\sum_{n=1}^\infty\frac{(\bb_i\cdot\nabla_\bk)^{2n-1}}{(2n-1)!}f(\bk)\sum_{m=1}^Nm^{2n-1}\tilde{c}_{m}^{(N)}\,.
\end{align}
By matching the coefficients of the first to the $(2n-1)$-th derivative of $f(\bk)$ of Eq.~\eqref{eq:fd_new_method} and Eq.~\eqref{eq:fd_new_method_taylor_exp}, similarly to what was done for the HOFD method under the constraint in Eq.~\eqref{eq:fd_naive}, we arrive at
\begin{equation} \label{eq:fd_taylor_exp_equality}
    \sum_{m=1}^Nm^{2n}\,\left(\frac{1}{m}\tilde{c}_{m}^{(N)}\right)=
    \begin{cases*}
    \frac{1}{2}\,\,\, \textrm{ if }\,\,\,n=1 \\
    0\,\,\, \textrm{ if }\,\,\,n=2,\,3,\,...,\,N \,.
    \end{cases*}
\end{equation}
Now, because Eq.~\eqref{eq:fd_taylor_exp_equality} is similar to Eqs.~\eqref{eq:eqs_1d}, \eqref{eq:eqs_1d_2}, and~\eqref{eq:vandermonde}, we can obtain $\tilde{c}_{m}^{(N)}$ using Eq.~\eqref{eq:cramers}:
\begin{equation} \label{eq:cramers2}
    \tilde{c}_{m}^{(N)} = \frac{1}{2m} \prod_{n \neq m}^{N} \frac{n^2}{n^2 - m^2}\,.
\end{equation}
Putting these $\tilde{c}_{m}^{(N)}$'s into Eq.~\eqref{eq:fd_new_method}, we have the FD expressions for $\bb_i\cdot\nabla_\bk f(\bk)$ $(i=1,\,\,2,\,\,...\,\,,\,\,N_\bb)$.
If we define an $N_\bb\times3$ matrix
\begin{equation}
    \label{eq:matrix_B}
    B=\begin{pmatrix}
        \bb_1^\textrm{T}\\
        \bb_2^\textrm{T}\\
        \vdots\\
        \bb_{N_\bb}^\textrm{T}
    \end{pmatrix}\,,
\end{equation}
in which we assumed that $\bb_i$ is a $3\times1$ matrix,
and
\begin{align} \label{eq:matrix_c}
    \textbf{c}=\sum_{m=1}^N \tilde{c}_m^{(N)}
    \begin{pmatrix}
      f(\bk+m\bb_1)- f(\bk-m\bb_1)\\
      f(\bk+m\bb_2) - f(\bk-m\bb_2) \\
      \vdots \\
      f(\bk+m\bb_{N_\bb}) - f(\bk-m\bb_{N_\bb})
    \end{pmatrix}\,,
\end{align}
which is an $N_\bb\times1$ matrix, we can find $\nabla_\bk f(\bk)$, a $3\times1$ matrix, from
\begin{align} \label{eq:Bx_eq_c}
B\,\nabla_\bk f(\bk)=\textbf{c}\,.
\end{align}
Since $N_\bb\ge3$ in three dimensions, the least squares solution for $\nabla_\bk f(\bk)$ is obtained by minimizing the error, $(B\,\nabla_\bk f(\bk)-\textbf{c})^\dagger (B\,\nabla_\bk f(\bk)-\textbf{c})$; the solution is
\begin{align} \label{eq:Bx_eq_c_sol}
\nabla_\bk f(\bk)=(B^\textrm{T}B)^{-1}B^\textrm{T}\textbf{c}\,,
\end{align}
where
\begin{equation}
    \label{eq:BTB}
    B^\textrm{T}B=\sum_{i=1}^{N_\bb} \bb_i\,\bb_i^\textrm{T}
\end{equation}
is a $3\times3$ matrix.
Note that if $N_\bb=3$, $B$ itself is a $3\times3$ matrix, and $\nabla_\bk f(\bk)=B^{-1}\textbf{c}$, which we can also obtain from Eq.~\eqref{eq:Bx_eq_c} by multiplying $B^{-1}$ on both sides.

This method has the following features:
(i) Conceptually, one first obtains the gradient of $f(\bk)$ along $\bb_i$, $\bb_i\cdot\nabla_\bk f(\bk)$ [Eq.~\eqref{eq:fd_new_method}], and then recovers $\nabla_\bk f(\bk)$ from these $\bb_i$-projected gradients.
(ii) The coefficients $\tilde{c}_{m}^{(N)}$ in Eq.~\eqref{eq:cramers2} do \textit{not} depend on the $\bb$ vector, making this method simple.
(iii) This method usually requires fewer $\bb$ vectors than the conventional FD method based on \Eq{eq:fd_naive}, which imposes additional, unnecessary constraints.
For example, only three non-coplanar $\bb$ vectors and their inversion pairs (or their $N$ multiples for the HOFD case) are needed if the system does not have any symmetry, which is a triclinic system, while the conventional method needs six $\bb$ vectors and their inversion pairs~\cite{1997Marzari}.

As an example for two dimensions, let us consider the situation of \Fig{fig:higher_order_search}(a), where the sampled $\bk$-points are from a grid $n_1 (1, 0) + n_2 (-0.25, 1)$ with integers $n_1$ and $n_2$. Then, one can choose $\bb_1 = (1, 0)$ and $\bb_2 = (-0.25, 1)$ to find the first-order FD value for $\nabla_\bk f(\bk)$ with four $\bb$ vectors according to Eq.~\eqref{eq:Bx_eq_c_sol} adapted for two dimensions, i.\,e.\,, $B$ is now a $2\times2$ matrix. To satisfy Eqs.~\eqref{eq:fd_naive} and~\eqref{eq:c_b_sum}, however, we need six $\bb$ vectors, as shown in \Fig{fig:higher_order_search}(a).

Still, we did not implement this method, as it requires a more involved modification of the existing FD implementation based on Eqs.~\eqref{eq:fd_naive} and~\eqref{eq:c_b_sum}. In addition, the overall computational cost does not depend strongly on the number of $\bb$ vectors used. However, if one wants to implement relevant functions from scratch, the method presented in this section is a good option, given the advantages discussed.

In practice, this scheme can be implemented by modifying the code that searches for $\bb$ vectors using \Eq{eq:fd_naive} (e.g., \textsc{Wannier90}) as follows.
We begin by generalizing the FD formula of \Eq{eq:fd_naive} to
\begin{equation} \label{eq:fd_general}
    \nabla_\bk f(\bk) \approx \sum_\bb \mb{g}_\bb f(\bkb) \,,
\end{equation}
where $\mb{g}_\bb$ are general vectors.
This parametrization is quite general and reduces to the conventional one~\cite{1997Marzari} [Eq.~\eqref{eq:c_b_sum}] or its HOFD variant [Eq.~\eqref{eq:fd_nearest_nd}] if we impose the constraint $\mb{g}_\bb=c_{\abs{\bb}}\bb$.
With this general parametrization, the first-order FD condition in the first line of \Eq{eq:fd_nearest_nd} becomes
\begin{equation} \label{eq:fd_nearest_general_1st}
    \sum_{\bb} g_\bb^{\alpha_1} b^{\alpha_2} = \delta_{\alpha_1 \alpha_2} \,.
\end{equation}
Using the $B$ matrix [\Eq{eq:matrix_B}] and similarly defining a $3\times N_\bb$ matrix
\begin{equation} \label{eq:matrix_Q}
    Q = \begin{pmatrix}
        \mb{g}_{\bb_1} &
        \mb{g}_{\bb_2} &
        \hdots &
        \mb{g}_{\bb_{N_\bb}}
    \end{pmatrix} \,,
\end{equation}
we can rewrite \Eq{eq:fd_nearest_general_1st} as
\begin{align} \label{eq:general_QB_identity}
    Q B = I \,,
\end{align}
where $I$ is a $3\times 3$ identity matrix.

Next, as in the usual case with $c_{\abs{\bb}}$ parametrization~\cite{1997Marzari, 2008MostofiWannier90}, we add shells of $\bb$ vectors one by one from the smallest $|\bb|$ to larger ones, until the $B$ matrix has full rank, i.\,e.\,, three.
This process guarantees the two conditions mentioned in the beginning of this section: the selected $\bb$ vectors are not all coplanar and the $\bb$ vectors are added in shells.
Once $B$ is of full rank, \Eq{eq:general_QB_identity} may have multiple solutions for $Q$.
We take the Moor--Penrose pseudoinverse solution, which is the least squares solution that minimizes the norm of $Q$:
\begin{align} \label{eq:new_method_Q_inverse}
    Q = B^\textrm{T} (B B^\textrm{T})^{-1} \,.
\end{align}
Substituting this $Q$ into \Eq{eq:fd_general}, we obtain the first-order version of \Eq{eq:Bx_eq_c_sol}.
Generalization to the HOFD case is done similarly to what was done in the multiples scheme under the constraint in \Eq{eq:fd_naive}, with the $\bb$ vectors chosen as multiples of the first-order ones.

In light of the generalized FD parameterization in Eq.~\eqref{eq:fd_general}, the result in Eq.~\eqref{eq:Bx_eq_c_sol} is equivalent to
\begin{align} \label{eq:new_method_qb}
    g_{\pm m\bb_i}^{\alpha_1}&= \pm \sum_{\beta_1}(B^\textrm{T}B)^{-1}_{\alpha_1 \beta_1}\,B^\textrm{T}_{\beta_1 i}\, \tilde{c}_m^{(N)}\nnnl
    &=\pm \sum_{\beta_1}(B^\textrm{T}B)^{-1}_{\alpha_1 \beta_1}\,b_i^{\beta_1}\, \tilde{c}_m^{(N)}\,,
\end{align}
or equivalently,
\begin{align} \label{eq:new_method_qb_vector}
    \mb{g}_{\pm m\bb_i}=\pm \left(\sum_{j=1}^{N_\bb} \bb_j \bb_j^\textrm{T}\right)^{-1}\,\bb_{i}\, \tilde{c}_m^{(N)}\,,
\end{align}
where we used Eq.~\eqref{eq:BTB}.
This is the final result to be used in computer codes (in place of $c_{\abs{\bb}}\,\bb$) together with Eq.~\eqref{eq:fd_general}.

\subsection{Symmetry property}
Next, we show that this method preserves the symmetry property of the gradient.
Suppose that $f(\bk)$ is a scalar function of $\bk$ having the symmetry of the system, i.\,e.\,, $f(S\bk)=f(\bk)$, where $S$ is the orthogonal transformation matrix for a symmetry operation of the system. We prove that our proposed method satisfies the desired symmetry property of the gradient, i.\,e.\,,
\begin{equation}
    \label{eq:grad_f_symm}
    (\nabla_\bk f)(S\bk) = S\,(\nabla_\bk f)(\bk)\,.
\end{equation}
Thus, using the generalized FD formula, we have to show that
\begin{align} \label{eq:grad_f_symm_FD1}
    (\nabla_\bk f)(S\bk) &\approx \sum_\bb \mb{g}_\bb f(S\bk+\bb)\nnnl
    &= \sum_\bb \mb{g}_\bb f(\bk+S^{-1}\bb)\nnnl
    &= \sum_\bb \mb{g}_{S\bb} f(\bk+\bb)
\end{align}
is equal to
\begin{align} \label{eq:grad_f_symm_FD2}
    S(\nabla_\bk f)(\bk) &\approx S\sum_\bb \mb{g}_\bb f(S\bk+\bb)\nnnl
    &= \sum_\bb S\mb{g}_\bb f(\bk+\bb)
\end{align}
or that
\begin{align} \label{eq:grad_f_symm_FD3}
    \mb{g}_{S\bb}= S\mb{g}_\bb\,.
\end{align}

This property can be verified for our proposed method as follows:
\begin{align} \label{eq:new_method_qSb}
    \mb{g}_{S(\pm m\bb_i)}&=\pm \left(\sum_j \bb_j \bb_j^\textrm{T}\right)^{-1}\,S\bb_{i}\, \tilde{c}_m^{(N)}\nnnl
    &=S\left[\pm S^\textrm{T}\left(\sum_j \bb_j \bb_j^\textrm{T}\right)^{-1}S\,\bb_{i}\, \tilde{c}_m^{(N)}\right]\nnnl
    &=S\left[\pm \left(S^\textrm{T}\sum_j \bb_j \bb_j^\textrm{T}S\right)^{-1}\,\bb_{i}\, \tilde{c}_m^{(N)}\right]\nnnl
    &=S\left[\pm \left(\sum_j S^\textrm{T}\bb_j (S^\textrm{T}\bb_j)^\textrm{T}\right)^{-1}\,\bb_{i}\, \tilde{c}_m^{(N)}\right]\nnnl
    &=S\left[\pm \left(\sum_j \bb_{j'} \bb_{j'}^\textrm{T}\right)^{-1}\,\bb_{i}\, \tilde{c}_m^{(N)}\right]\nnnl
    &=S\left[\pm \left(\sum_j \bb_{j} \bb_{j}^\textrm{T}\right)^{-1}\,\bb_{i}\, \tilde{c}_m^{(N)}\right]\nnnl
    &=S\mb{g}_{\pm m\bb_i}\,.
\end{align}
In the fifth equality, we used the property that $S\bb_j=\bb_{j'}$ for some $j'$ because we included all the $\bb$ vectors on the same shell.

\subsection{Combining with the TEFD method}

This proposed method can be straightforwardly combined with the TEFD formalism developed in Secs.~\ref{sec:position} and~\ref{sec:composite}.
One simply has to replace $c_{\abs{\bb}}\,\bb$ with $\mb{g}_\bb$ [Eq.~\eqref{eq:new_method_qb_vector}] in the final expressions therefrom, for example, the equations in the TEFD column of Tab.~\ref{tab:summary}.

The symmetry property in Eq.~\eqref{eq:new_method_qSb}, when combined with the TEFD method, satisfies the symmetry relation of the relevant matrix elements. For example, if $c_{\abs{\bb}}\,(S\bb)=c_{\abs{S\bb}}\,(S\bb)$ in Eq.~\eqref{eq:sym_r_TEFD} is replaced with $\mb{g}_{S\bb}=S\mb{g}_\bb$ [Eq.~\eqref{eq:new_method_qSb}], the symmetry relation of the position matrix element is still satisfied.

\section{Conclusion}
\label{sec:conclusion}

In conclusion, we proposed two key improvements to the finite-difference calculation of the position and composite matrix elements used in Wannier interpolation of $\bk$-derivatives and quantum geometric quantities.
First, we demonstrated that satisfying translational equivariance is crucial for the reliability and accuracy of finite-difference methods and developed equivariant expressions for the diagonal and off-diagonal matrix elements of the position and composite operators.
Second, we introduced higher-order finite-difference schemes for computing these matrix elements.
By applying these methods to real-material calculations, we showed that they significantly improve the convergence and accuracy of various quantities, namely the Wannier centers and spreads, electric polarization, position matrix elements, orbital magnetization, and spin Hall conductivity.
Finally, we proposed a vector-coefficient scheme that can reduce the number of FD vectors while maintaining or even improving accuracy.
Our work offers a thorough analysis of finite-difference approximations for the position operator and $\bk$-space derivatives.

After a preliminary version of this work was presented~\cite{Lihm2022talk, Ghim2022talk}, our TEFD and HOFD methods have been implemented in open-source software packages \textsc{Wannier90}~\cite{2008MostofiWannier90, 2020PizziWannier90}, \textsc{WannierBerri}~\cite{tsirkin2021WB}, and \textsc{EPW}~\cite{Ponce2022, Lee2023EPW}.
Moreover, thanks to the versatility of our approach that can be applied to any composite operators, the TEFD method has already been applied to formulate the Wannier interpolation of spatially dispersive optical conductivity by \citet{Urru2025} and shown to improve the symmetry of the shift current response~\cite{WBerriIssue}.
Our new approaches introduce minimal additional code complexity or computational overhead relative to existing implementations, while significantly improving the accuracy and efficiency of Wannier interpolation.
We anticipate that the proposed methods will be widely adopted across the Wannier-function-based software ecosystem~\cite{2024MarrazzoWannier}.

\medskip
\textit{Note added.}---Recently, a related study appeared on the arXiv~\cite{Thummler2026}.

\acknowledgments

We thank Sinisa Coh, Nicola Marzari, Arash Mostofi, Samuel Ponc\'e, \'Oscar Pozo Oca\~na, Daniel S\'anchez Portal, Junfeng Qiao, Ivo Souza, Massimiliano Stengel, Stepan Tsirkin, and Jonathan Yates for helpful discussion.
We especially thank Ivo Souza for comparing notes on the translationally equivariant matrix elements for orbital magnetization.
This work was supported by the Korean NRF No-2023R1A2C1007297.
J.-M.L. was partially supported by the Fonds de la Recherche Scientifique - FNRS under Grants number T.0183.23 (PDR) and T.W011.23 (PDR-WEAVE).
Computational resources have been provided by KISTI (KSC-2023-CRE-0533) and the Consortium des \'Equipements de Calcul Intensif, funded by the FRS-FNRS under Grant No.2.5020.11.
The Flatiron Institute is a division of the Simons Foundation.

\medskip
\begin{center}{\small\bfseries\MakeUppercase Data and Code Availability}\end{center}

The code developed in this study is available in the Wannier90 and WannierBerri repositories.
All inputs and outputs and the scripts to produce the figures for this work will be made available on the Materials Cloud Archive upon publication.





\appendix

\section{Size consistency of FD approximations} \label{sec:size_consistency}

In this section, we discuss whether the different FD approximations are size consistent.
By size consistency, we mean that the unit cell with an $N_k \times N_k \times N_k$ Brillouin zone sampling should be equivalent to an $N_k \times N_k \times N_k$ supercell with a $\Gamma$-only sampling~\cite{2006Stengel}.
We show that the S\&S and TEFD methods are size consistent, while the S-FD and M\&V methods are not.

The WFs of a supercell can be indexed as $I = (i, \bR_I)$, where $\bR_I$ is a unit cell lattice vector.
The supercell WFs have a one-to-one mapping to the unit cell ones:
\begin{equation}
    \ket{w^{\rm sc}_{I}}
    = \ket{w_{i\bR_I}}
    = \frac{1}{\sqrt{N_\bk}} \sum_\bk \ket{\psi^{\rm (W)}_\ik} e^{-i \bk\cdot \bR_I} \,.
\end{equation}
By using the first line of \Eq{eq:M_ijR} while noting that the supercell case has no $\bR$ summation, we find
\begin{align} \label{eq:sc_M}
    M^{{\rm sc}\,(\mb{0},\mb{b})}_{IJ}
    &= \mel{w^{\rm sc}_{I}}{e^{-i\bb\cdot\opbr}}{w^{\rm sc}_{J}}
    \nnnl
    &= \frac{1}{N_\bk} \sum_{\bk\bk'} \mel{\psi^{\rm (W)}_\ik}{e^{-i\bb\cdot\opbr}}{\psi^{\rm (W)}_{j\bk'}}
    e^{i \bk\cdot \bR_I} e^{-i \bk'\cdot \bR_J}
    \nnnl
    &= \frac{1}{N_\bk} \sum_{\bk} \mel{\psi^{\rm (W)}_\ik}{e^{-i\bb\cdot\opbr}}{\psi^{\rm (W)}_\jkb}
    e^{i \bk\cdot \bR_I} e^{-i (\bk+\bb)\cdot \bR_J}
    \nnnl
    &= \frac{1}{N_\bk} e^{-i \bb \cdot \bR_J} \sum_{\bk} M^\kb_{ij} e^{i \bk \cdot (\bR_I - \bR_J)} \,.
\end{align}
In the third equality, we used the fact that the matrix element vanishes unless $\bk' = \bk + \bb$.

We now show the size consistency and the lack thereof for the FD approximations for the Wannier centers and position matrix elements.
We begin with the S\&S formula for the Wannier center [\Eq{eq:center_SS}], which was shown to be size consistent in the original paper of \citet{2006Stengel}.
Concretely, the supercell Wannier centers computed using the S\&S method are identical to the unit cell ones.
We can directly see it by substituting \Eq{eq:sc_M} into \Eq{eq:center_SS}:
\begin{align} \label{eq:sc_center_SS}
    \br^{\rm sc,\, \rmSS}_{I}
    &= - \sum_\bb c_{\abs{\bb}} \bb \Im \ln M^{{\rm sc}\,(\mb{0},\bb)}_{II}
    \nnnl
    &= - \sum_\bb c_{\abs{\bb}} \bb \Im \ln \biggl( \frac{1}{N_\bk} e^{-i \bb \cdot \bR_I} \sum_{\bk} M^\kb_{ii} \biggr)
    \nnnl
    &= \br^{\rmSS}_{i} + \sum_\bb c_{\abs{\bb}} \bb (\bb \cdot \bR_I)
    \nnnl
    &= \br^{\rmSS}_{i} + \bR_I \,.
\end{align}
In the last equality, we used the FD condition, \Eq{eq:c_b_sum}.
In contrast, for the M\&V formula, the supercell and unit-cell results are inconsistent.
In fact, the M\&V formula applied to the supercell reduces to the S\&S result on the unit cell; plugging \Eq{eq:sc_M} into \Eq{eq:center_MV}, we find
\begin{align} \label{eq:sc_center_MV}
    \br_{I}^{\rm sc,\, \rmMV}
    &= -\sum_{\bb} c_{\abs{\bb}} \bb \Im \ln M^{{\rm sc}\,(\mb{0},\bb)}_{II}
    \nnnl
    &= \br^{\rmSS}_{i} + \bR_I
    \nnnl
    &\neq \br^{\rmMV}_{i} + \bR_I
    \,.
\end{align}

Next, we consider the full position matrix.
For the TEFD case [\Eq{eq:tr_r_fd_trinv}], we find
\begin{align} \label{eq:sc_r_tefd}
    &\quad \br_{IJ}^{{\rm sc},\, \rmTEFD}
    \nnnl
    &= \br^{\rm sc}_I \delta_{IJ}
    + i \sum_{\bb} c_{\abs{\bb}} \bb \, e^{i\bb\cdot \frac{\br^{\rm sc}_I + \br^{\rm sc}_J}{2}} M_{IJ}^{{\rm sc}\,(\mb{0},\bb)}
    \nnnl
    &= \br^{\rm sc}_I \delta_{IJ}\nnnl
    &+ \frac{i}{N_\bk} \sum_{\bk\bb} c_{\abs{\bb}} \bb \, e^{i\bb\cdot \frac{\br_i + \br_j + \bR_I + \bR_J}{2}} e^{-i \bb \cdot \bR_J} e^{i \bk \cdot (\bR_I - \bR_J)} M_{ij}^\kb
    \nnnl
    &= \br^{\rm sc}_I \delta_{IJ}\nnnl
    &+ \frac{i}{N_\bk} \sum_{\bk\bb} c_{\abs{\bb}} \bb \, e^{i\bb\cdot \frac{\br_i + \br_j + \bR_J - \bR_I}{2}} e^{-i (\bk+\bb) \cdot (\bR_J - \bR_I)} M_{ij}^\kb
    \nnnl
    &= (\br_i + \bR_I) \delta_{ij} \delta_{\bR_J \bR_I}
    + \br_{ij\bR_J-\bR_I}^{\rmTEFD} \,.
\end{align}
This is the expected result respecting size-consistency, where the supercell position matrix elements are equivalent to the unit cell ones plus an additional diagonal term due to Wannier center shifts.

In contrast, for the S-FD formula [\Eq{eq:r_fd}], the size-consistency is broken.
For the Wannier centers, we find
\begin{align} \label{eq:sc_r_fd}
    \br_{I}^{{\rm sc},\, \rmSFD}
    &= i \sum_{\bb} c_{\abs{\bb}} \bb \, M_{II}^{{\rm sc}\,(\mb{0},\bb)}
    \nnnl
    &= \frac{i}{N_\bk} \sum_{\bk\bb} c_{\abs{\bb}} \bb \, M^\kb_{ii} e^{-i \bb \cdot \bR_I}
    \nnnl
    &\neq \frac{i}{N_\bk} \sum_{\bk\bb} c_{\abs{\bb}} \bb \, M^\kb_{ii} + \bR_I
    \nnnl
    &= \br_{i}^{\rmSFD} + \bR_I \,.
\end{align}
The supercell and unit-cell Wannier centers agree at $\bR_I = \mb{0}$, but not at $\bR_I \neq \mb{0}$.
For the off-diagonal case $I \neq J$, we find
\begin{align} \label{eq:sc_r_od_fd}
    \br_{IJ}^{{\rm sc},\, \rmSFD}
    &= i \sum_{\bb} c_{\abs{\bb}} \bb \, M_{IJ}^{{\rm sc}\,(\mb{0},\bb)}
    \nnnl
    &= \frac{i}{N_\bk} \sum_{\bk\bb} c_{\abs{\bb}} \bb \, M^\kb_{ij} e^{i \bk \cdot (\bR_I - \bR_J)} e^{-i \bb \cdot \bR_J}
    \nnnl
    &\neq \frac{i}{N_\bk} \sum_{\bk\bb} c_{\abs{\bb}} \bb \, M^\kb_{ij} e^{-i \bk \cdot (\bR_J - \bR_I)}
    \nnnl
    &= \br_{ij\bR_J-\bR_I}^{\rmSFD} \,.
\end{align}

\section{Computational details} \label{app:computational}

\begin{table*}
\begin{tabular}{c|ccc|ccccc}
\hline
Material & $E_{\rm cut}$ (Ry) & $N_\bk^{\rm DFT}$ & SOC
& $N^{\rm band}$ & $N^{\rm Wannier}$
& Initial guess & $E^{\rm froz}$ (eV) & $E^{\rm dis}$ (eV)
\\ \hline
GeS   & 70  & $12^2$ & \XSolidBrush
& 22 & 16
& Ge: $s,p$, S: $s,p$ & 2.5 & 6.0
\\
Graphene & 80  & $12^2$ & \XSolidBrush
& 15 & 2
& C: $p_z$ & [$-2.8$, 2.7] & 13.7
\\
Fe   & 120  & $16^3$ & \Checkmark
& 50 & 26
& Fe: semicore-$s,p$, $sp^3d^2$, $t_{2g}$ & 12.3 & 68.0
\\
CrO$_2$ & 100 & $6^3$ & \Checkmark
& 100 & 68
& Cr: $s,p,d$, O: $s,p$ & 4.4 & 26.6
\\
GaAs  & 80 & $12^3$ & \Checkmark
& 36 & 18
& Ga: $s,p$, As: $s,p$, Interstitial: $s$ & 5.6 & 25.3
\\
Si  & 70 & $14^3$ & \XSolidBrush
& 4 & 4
& Bond-centered $s$ & -- & --
\\
KNbO$_3$ & 70 & $6^3$ & \XSolidBrush
& 15 & 15
& K: $p$, O: $s,p$ & -- & --
\\
\hline
\end{tabular}
\caption{
    Computational parameters used in this work: kinetic energy cutoff $E_{\rm cut}$,
    size of the $\bk$-point grid for the density functional theory (DFT) calculation $N_\bk^{\rm DFT}$,
    inclusion of the spin-orbit coupling (SOC),
    number of bands included for Wannierization $N^{\rm band}$,
    number of the WFs $N^{\rm Wannier}$,
    initial guesses for constructing the Wannier functions,
    upper bound (and lower bound if present) of the frozen (inner) window $E^{\rm froz}$, and
    upper bound of the disentanglement (outer) window $E^{\rm dis}$.
    The frozen and disentanglement windows are defined with respect to the conduction band minimum energy for insulators and the Fermi level for metals.
    Dashes denote that the corresponding window was not set.
}
\label{tab:comp_params}
\end{table*}

We used \qe~\cite{giannozzi2009quantum,giannozzi2017advanced} to perform DFT calculations and \textsc{Wannier90}~\cite{2008MostofiWannier90, 2020PizziWannier90} to construct the WFs.
Wannier interpolation of the position and composite matrix elements, as well as the calculation of the orbital magnetization and spin Hall conductivity, are done with \textsc{Wannier90}~\cite{2008MostofiWannier90, 2020PizziWannier90} and \textsc{WannierBerri}~\cite{tsirkin2021WB}.
We also used the \textsc{Wannier.jl}~\cite{Qiao2023, Wannierjl} package for testing the implementation of TEFD.
The spin Hall conductivity was computed using the method of Ref.~\cite{2019RyooSHC}.
We modified the \textsc{Wannier90} and \textsc{WannierBerri} codes to implement the TEFD and HOFD methods developed in this work.
These modifications are available in the develop branch of the respective programs and will be included in the next release.

We used pseudopotentials from \textsc{PseudoDojo} (v4)~\cite{2018vanSettenPseudodojo} in the Perdew--Burke--Ernzerhof generalized-gradient approximation~\cite{1996PerdewPBE}.
For GeS, we used the relaxed structures from Ref.~\cite{Rangel2017GeS}.
For CrO$_2$, we relaxed the structure and lattice parameters until the forces and stresses were under $10^{-3}~\mathrm{Ry/bohr}$ and 0.5~kbar, respectively.
For other cases, we used fixed lattice constants of 2.46~\AA\ (graphene), 5.42~\AA\ (Fe), 5.65~\AA\ (GaAs), 5.43~\AA\ (Si), and $a=4.00~\text{\AA}$, $c/a = 1.0165~\text{\AA}$ (KNbO$_3$).
For all calculations, we used disentanglement-only WFs without performing maximal localization to make sure that the WFs maintain the orbital character of the initial guess.
The size of the coarse $\bk$-point grid for Wannierization is specified in each figure.
For the calculation of orbital magnetization, we used a $120^3$ fine $\bk$ grid for both bcc Fe and CrO$_2$.
For the spin Hall conductivity of GaAs, we used a $150^3$ fine $\bk$-point grid.
Other computational parameters are summarized in Tab.~\ref{tab:comp_params}.

\section{Symmetry of the TEFD method}
\label{sec:symmetry}

In this section, we show that the TEFD formula for the position operator respects the crystal symmetries, while the S-FD formula does not.
Consider a symmetry operation of the system described by a rotation matrix $S$ and a fractional translation vector $\mb{f}$, which transforms the position operator as $\opbr \to S\opbr + \mb{f}$.
In general, one WF may transform as a linear combination of other WFs under a symmetry operation, and such a transformation is described by a unitary matrix~\cite{2013SakumaSymmetryadapted}.
Although the following proof can be straightforwardly extended to the general case, for simplicity, we consider the case where a WF transforms to another WF (not to a linear combination of multiple WFs):
\begin{equation} \label{eq:sym_wf}
    w_{i\mb{0}}(\br) = w_{i'\bR_i}(S\br + \mb{f}) \,.
\end{equation}
In momentum space, the corresponding cell-periodic parts of the Bloch functions transform as
\begin{align} \label{eq:sym_u}
    &\quad u_{i\bk}(\br)
    \nnnl
    &= \sum_\bR w_{i\bR}(\br) e^{i\bk\cdot(\bR - \br)}
    \nnnl
    &= \sum_\bR w_{i\mb{0}}(\br - \bR) e^{i\bk\cdot(\bR - \br)}
    \nnnl
    &= \sum_\bR w_{i'\bR_i}(S\br + \mb{f} - S \bR) e^{i\bk\cdot(\bR - \br)}
    \nnnl
    &= \sum_\bR w_{i' S\bR + \bR_i}(S\br + \mb{f}) e^{i(S\bk)\cdot(S\bR + \bR_i - S\br - \mb{f})} e^{i(S\bk)\cdot (\mb{f} - \bR_i)}
    \nnnl
    &= u_{i' S\bk}(S\br + \mb{f}) \, e^{i(S\bk)\cdot (\mb{f} - \bR_i)} \,.
\end{align}
In the fourth equality, we used $e^{i(S\bk)\cdot(S\bR)} = e^{i\bk\cdot\bR}$.

Next, we derive the symmetry relation for the $M$ matrix.
We start from its definition in \Eq{eq:M_def} and substitute \Eq{eq:sym_wf} into it:
\begin{align} \label{eq:sym_M}
    M_{ij}^\kb
    &= \nint \dd\br \, [u^\rmW_\ik(\br)]^* \, u^\rmW_\jkb(\br)
    \nnnl
    &= \nint \dd\br \,
    e^{-i(S\bk)\cdot(\mb{f} - \bR_i)} \,
    e^{i(S\bk+S\bb)\cdot(\mb{f} - \bR_j)}
    \nnnl
    &\quad \times
    [u_{i'S\bk}(S\br + \mb{f})]^* \,
    u^\rmW_{j'S\bk+S\bb}(S\br + \mb{f})
    \nnnl
    &= M_{i' j'}^{(S\bk,S\bb)} \,
    e^{i(S\bk)\cdot(\bR_i - \bR_j)}
    e^{i(S\bb)\cdot(\mb{f} - \bR_j)} \,.
\end{align}

Now, let us examine the symmetry of the position matrix element.
In an exact calculation, the desired symmetry relation is
\begin{align} \label{eq:sym_r_exact}
    &\quad S\br_\ijR
    \nnnl
    &= \nint \dd\br \, S\br \, w_{i\mb{0}}^*(\br) \, w_{j\bR}(\br)
    \nnnl
    &= \nint \dd\br \, S\br \, w_{i\mb{0}}^*(\br) \, w_{j\mb{0}}(\br-\bR)
    \nnnl
    &= \nint \dd\br \, S\br \, w_{i'\bR_i}^*(S\br + \mb{f}) \, w_{j' \bR_j}(S(\br-\bR) + \mb{f})
    \nnnl
    &= \nint \dd\br \, S\br \, w_{i'\bR_i}^*(S\br + \mb{f}) \, w_{j' S\bR + \bR_j}(S\br + \mb{f})
    \nnnl
    &= \nint \dd\br' \, (\br'-\mb{f}) \, w_{i'\bR_i}^*(\br') \, w_{j' S\bR + \bR_j}(\br')
    \nnnl
    &= \nint \dd\br' \, (\br'-\mb{f}) \, w_{i'\mb{0}}^*(\br' - \bR_i) \, w_{j' S\bR + \bR_j - \bR_i}(\br' - \bR_i)
    \nnnl
    &= \nint \dd\br' \, (\br'+\bR_i-\mb{f}) \, w_{i'\mb{0}}^*(\br') \, w_{j' S\bR + \bR_j - \bR_i}(\br')
    \nnnl
    &= \br_{i'j'S\bR+\bR_j-\bR_i}
    + \delta_{ij} \delta_{\bR\mb{0}} \, (\bR_i - \mb{f}) \,.
\end{align}
In the last equality, we used the fact that for the two transformed WFs to be identical, the two original WFs must also be identical, i.e.,
\begin{equation}
    \delta_{i'j'} \delta_{S\bR + \bR_j- \bR_i, \mb{0}}
    = \delta_{ij} \delta_{\bR\mb{0}} \,.
\end{equation}
Equivalently, the symmetry relation can be expressed as
\begin{equation} \label{eq:sym_r_exact_final}
    \br_{i'j'S\bR+\bR_j-\bR_i}
    = S \br_\ijR
    + \delta_{ij} \delta_{\bR\mb{0}} \,(\mb{f} - \bR_i) \,.
\end{equation}
The diagonal part of this equation gives the relation of the centers of the original and symmetry-transformed WFs:
\begin{equation} \label{eq:sym_center}
    \mb{r}_{i'} = S \br_i + \mb{f} - \bR_i \,.
\end{equation}

We now show that the TEFD formula satisfies the symmetry relation \Eq{eq:sym_r_exact_final}, while the S-FD formula does not.
Let us begin with the TEFD formula.
The midpoint $\bar{\br}_{\ijR}$ defined in \Eq{eq:tr_r_midpoint_def} satisfies
\begin{align} \label{eq:sym_r_midpoint}
    S \bar{\br}_{\ijR}
    &= \frac{S\br_i + S \br_j  + S \bR}{2} \nnnl
    &= \frac{\br_{i'} + \br_{j'} - 2\mb{f} + \bR_i + \bR_j + S \bR}{2} \nnnl
    &= \bar{\br}_{i'j'S\bR + \bR_j - \bR_i} + \bR_i - \mb{f} \,.
\end{align}
Substituting \Eq{eq:sym_M} into \Eq{eq:tr_r_fd_trinv} and changing variables $\bk$ and $\bb$ to $S\bk$ and $S\bb$, we have
\begin{widetext}
\begin{align} \label{eq:sym_r_TEFD}
    &\quad \br_{i'j'S\bR+\bR_j-\bR_i}^\rmTEFD
    \nnnl
    &= \br_{i'} \delta_{ij} \delta_{\bR\mb{0}}
    + \frac{i}{N_\bk} \sum_{\bb} c_{\abs{\bb}} (S \bb) \, e^{i(S\bb)\cdot \bar{\br}_{i'j'S\bR + \bR_j - \bR_i}} \sum_{\bk} e^{-i(S\bk + S\bb)\cdot(S\bR + \bR_j - \bR_i)}
    M_{ij}^{(\bk,\bb)}
    e^{-i(S\bk)\cdot(\bR_i - \bR_j)}
    e^{-i(S\bb)\cdot(\mb{f} - \bR_j)}
    \nnnl
    &= \br_{i'} \delta_{ij} \delta_{\bR\mb{0}}
    + \frac{i}{N_\bk} \sum_{\bb} c_{\abs{\bb}} (S \bb) \,
    e^{i \bb \cdot \bar{\br}_{\ijR}}
    \sum_{\bk}
    e^{-i (\bk + \bb) \cdot \bR}
    M_{ij}^{(\bk,\bb)}
    \nnnl
    &= (S\br_i + \mb{f} - \bR_i) \delta_{ij} \delta_{\bR\mb{0}}
    + \frac{i}{N_\bk} \sum_{\bb} c_{\abs{\bb}} (S \bb) \,
    e^{i \bb \cdot \bar{\br}_{\ijR}}
    \sum_{\bk}
    e^{-i (\bk + \bb) \cdot \bR}
    M_{ij}^{(\bk,\bb)}
    \nnnl
    &= S \br_\ijR^\rmTEFD
    + (\mb{f} - \bR_i) \delta_{ij} \delta_{\bR\mb{0}}
    \,.
\end{align}
\end{widetext}
In the first equality, we used $c_{\abs{\bb}} = c_{\abs{S\bb}}$.
In the second equality, we used \Eq{eq:sym_r_midpoint} and $e^{i(S\bk)\cdot(S\bR)} = e^{i\bk\cdot\bR}$.
In the third equality, we used \Eq{eq:sym_center} to rewrite $\br_{i'}$.
In the last equality, we used \Eq{eq:tr_r_fd_trinv}.
We thus recover the desired symmetry relation in \Eq{eq:sym_r_exact_final}.

Next, we consider the S-FD formula [\Eq{eq:r_fd}].
Following the same procedure as above, we have
\begin{align} \label{eq:sym_r_fd}
    &\quad \br_{i'j'S\bR+\bR_j - \bR_i}^{\rmSFD}
    \nnnl
    &= \frac{i}{N_\bk} \sum_{\bk\bb} c_{\abs{\bb}} (S \bb) \,
    e^{-i(S\bk)\cdot(S\bR + \bR_j - \bR_i)}
    \nnnl
    &\qquad\qquad \times
    M_{ij}^{(\bk,\bb)}
    e^{-i(S\bk)\cdot(\bR_i - \bR_j)}
    e^{-i(S\bb)\cdot(\mb{f} - \bR_j)}
    \nnnl
    &= S\left[\frac{i}{N_\bk} \sum_{\bk\bb} c_{\abs{\bb}} \bb \,
    e^{-i \bk\cdot \bR}
    M_{ij}^{(\bk,\bb)}
    e^{-i(S\bb)\cdot(\mb{f} - \bR_j)}\right]
    \nnnl
    &\neq S\left[\frac{i}{N_\bk} \sum_{\bk\bb} c_{\abs{\bb}} \bb \,
    e^{-i \bk\cdot \bR}
    M_{ij}^{(\bk,\bb)}
    \right]+(\mb{f} - \bR_i) \delta_{ij} \delta_{\bR\mb{0}}
    \nnnl
    &= S\br_{ij\bR}^{\rmSFD}+(\mb{f} - \bR_i) \delta_{ij} \delta_{\bR\mb{0}}
    \,.
\end{align}
Thus, the S-FD formula does not respect the crystal symmetries [\Eq{eq:sym_r_exact_final}].

\section{Accurate computation of spin-velocity matrix elements}
\label{sec:spin_vel}

\begin{table*}[]
\begin{tabular}{c|c|c|c}
                    & Formula & Matrix elements & Approximations \\ \hline
\citet{2018QiaoSHC} & Eqs.~(30),~(31), and~(32) of Ref.~\cite{2018QiaoSHC} & $H$, $r$, $s$ & \makecell{Completeness of calculated Bloch states \\ Can be numerically unstable}
\\ \hline
\citet{2019RyooSHC} & Eq.~(26) of Ref.~\cite{2019RyooSHC} & $H$, $r$, $s$, $sr$, $sHr$ & Dot equality in Eq.~\eqref{eq:invalid_dot_equality}
\\ \hline
This work           & Eqs.~\eqref{eq:revised_dot} and~\eqref{eq:dot2_revised} & $H$, $r$, $s$,  $sr$, $sHr$ & None
\end{tabular}
\caption{
    A comparison of three methods for the Wannier interpolation of spin-velocity matrix elements.
    The second column lists the real-space matrix elements needed to be computed.
    The third column details the approximation used in each method.
}
\label{tab:spin_vel_methods}
\end{table*}

Two different methods for the Wannier interpolation of spin-velocity have been proposed in Refs.~\cite{2018QiaoSHC} and \cite{2019RyooSHC}.
Yet, no work has theoretically analyzed and compared the two methods. In this section, we analyze the two methods and also propose a third, more accurate interpolation scheme.

Let us first review the steps for Wannier interpolation.
Suppose that we want to Wannier interpolate the Bloch matrix element $\mel{u_{m\bk}}{\hat{O}_\bk}{u_{n\bk}}$ where $\hat{O}_\bk = e^{-i\bk\cdot\opbr}\hat{O}e^{i\bk\cdot\opbr}$ and $\hat{O}$ is a periodic operator.
The first step is to compute the matrix elements on the coarse \textit{ab initio} grid of $\bk$ points.
Then, one finds gauge transform matrices $V_\bk$ that convert the wavefunctions from the rapidly varying eigenstate gauge to the Wannier gauge, which is smooth in $\bk$.
We write the transformation as
\begin{equation} \label{eq:uW_V_u0}
    \ket{u^{(\textrm{W})}_{i\bk}} = \sum_n  \ket{u^{(0)}_{n\bk}} \,V_{ni\bk}\,,
\end{equation}
where $\ket{u_{n\bk}^{(0)}}$'s are \textit{ab initio} eigenstates.
We use the superscript (0) to denote DFT wavefunctions to distinguish them from Wannier-interpolated energy eigenstates, which we denote without any superscripts.
$V_\bk$ is an $N^{\rm band}(\bk) \times N_{\rm W}$ matrix where $N^{\rm band}(\bk)$ is the number of bands at a given $\bk$ point considered in the Wannierization, and $N_{\rm W}$ is the number of WFs.
The gauge matrix is usually determined by a projection to atomic initial guess functions, often followed by an optimization to minimize the real-space spreads of the WFs.
The matrix elements in the Wannier gauge read
\begin{equation}
    \mel{u_{i\bk}^\rmW}{\hat{O}_\bk}{u_{j\bk}^\rmW} = \sum_{mn} V^\dagger_{im\bk} \mel{u_{m\bk}^{(0)}}{\hat{O}_\bk}{u_{n\bk}^{(0)}} V_{nj\bk}\,.
\end{equation}

Then, in the smooth Wannier gauge, one Fourier interpolates $\mel{u_{i\bk}^\rmW}{\hat{O}_\bk}{u_{j\bk}^\rmW}$ from the coarse \textit{ab initio} grid to the fine grid.
Finally, one converts the matrix elements back to the eigenstate gauge, where the transformation matrix is obtained by diagonalizing the Fourier interpolated Hamiltonian matrix $H_{ij\bk}^\rmW = \mel{u_{i\bk}^\rmW}{\hat{H}_\bk}{u_{j\bk}^\rmW}$:
\begin{equation}
    \sum_{ij} U_{mi\bk} H_{ij\bk}^\rmW U^\dagger_{jn\bk} = \varepsilon_{m\bk} \delta_{mn} \,,
\end{equation}
and
\begin{equation} \label{eq:spin_vel_H_eigen}
    \ket{u_{m\bk}} = \sum_{i} \ket{u_{i\bk}^\rmW} U^\dagger_{im\bk} \, .
\end{equation}
Here, $\varepsilon_{m\bk}$ is the interpolated eigenvalue, and $U_\bk$ an $N_{\rm W}\times N_{\rm W}$ unitary matrix.

When $N^{\rm band} > N_{\rm W}$, only a subset of Wannier interpolated bands, those inside the frozen (inner) energy window $\mathcal{W}_{\rm froz}$, interpolate \textit{ab initio} bands.
In other words, $\ket{u_{m\bk}}$ is an eigenstate of the \textit{ab initio} Hamiltonian $\hat{H}_\bk$ only if $\mk \in \mathcal{W}_{\rm froz}$:
\begin{equation}
    \hat{H}_\bk \ket{u_{m\bk}} = \varepsilon_{m\bk} \ket{u_{m\bk}} \ \text{only if}\ \mk \in \mathcal{W}_{\rm froz} \,.
\end{equation}

We now examine the specific case of the Wannier interpolation of the spin-velocity matrix, which is needed to calculate the spin Hall conductivity~\cite{Guo2005SHC}.
The spin-velocity matrix is defined as
\begin{equation} \label{eq:spin_vel_def}
  j^{\alpha\beta}_{mn\bk} = \tfrac{1}{2} \mel{u_\mk}{\{ \hat{s}^\alpha, \hat{v}^{\beta}_\bk\}}{u_\nk} \,,
\end{equation}
where $\hat{v}^\beta_{\bk} = \frac{1}{\hbar} \partial_\beta \hat{H}_{\bk}$ is the velocity operator in the $\beta$ direction.
It suffices to compute the matrix elements of $\hat{s}^\alpha\hat{v}^\beta_\bk$ as $\hat{v}^\beta_\bk\hat{s}^\alpha$ is its hermitian conjugate.
The matrix element can be faithfully interpolated only for bands inside the frozen window.
Hereafter in this section, we always assume that the states $\mk$ and $\nk$ are within the frozen window.

Let us first examine the method of \citet{2018QiaoSHC}.
This method aims to rewrite the spin-velocity in terms of the spin and position matrix elements and their derivatives, but nothing more, so that one does not need any additional \textit{ab initio} matrix elements to be computed.
In the Hamiltonian gauge, the spin-velocity in the inner window reads
\begin{align} \label{eq:spin_vel_Qiao_1}
   &\quad  \hbar \mel{u_{m\bk}}{\hat{s}^\alpha \hat{v}_{\beta,\bk}}{u_{n\bk}}
   \nnnl
   &= \mel{u_{m\bk}}{\hat{s}^\alpha (\partial_\beta  \hat{H}_\bk)}{u_{n\bk}}
   \nnnl
   &= \partial_\beta \mel{u_{m\bk}}{\hat{s}^\alpha \hat{H}_\bk}{u_{n\bk}} - \mel{\partial_\beta u_{m\bk}}{\hat{s}^\alpha \hat{H}_\bk}{u_{n\bk}}
   \nnnl &\quad
   - \mel{u_{m\bk}}{\hat{s}^\alpha \hat{H}_\bk}{\partial_\beta u_{n\bk}}
   \nnnl &=
   \partial_\beta \left ( \varepsilon_{n\bk}\mel{u_{m\bk}}{\hat{s}^\alpha}{u_{n\bk}} \right ) - \varepsilon_{n\bk} \mel{\partial_\beta u_{m\bk}}{\hat{s}^\alpha}{u_{n\bk}}
   \nnnl &\quad
   - \mel{u_{m\bk}}{\hat{s}^\alpha \hat{H}_\bk}{\partial_\beta u_{n\bk}}
   \nnnl &=
   \partial_\beta \varepsilon_{n\bk} \mel{u_{m\bk}}{\hat{s}^\alpha}{u_{n\bk}} + \varepsilon_{n\bk} \mel{u_{m\bk}}{\hat{s}^\alpha}{\partial_\beta u_{n\bk}}
   \nnnl &\quad
   - \mel{u_{m\bk}}{\hat{s}^\alpha \hat{H}_\bk}{\partial_\beta u_{n\bk}} \,.
\end{align}
Then, quantities involving the wavefunction derivatives are approximated using the FD method, which requires \textit{ab initio} matrix elements $\mel{u_{m\bk}^{(0)}}{\hat{s}^\gamma}{u_{n\bk+\bb}^{(0)}}$ and $\mel{u_{m\bk}^{(0)}}{\hat{s}^\gamma \hat{H}_\bk}{u_{n\bk+\bb}^{(0)}}$.
Finally, in the second equalities of Eqs.~(50) and (51) of Ref.~\cite{2018QiaoSHC}, one uses the completeness of the \textit{ab initio} eigenstates and rewrites these terms as
\begin{align} \label{eq:Qiao_2}
    \mel{u_{m\bk}^{(0)}}{\hat{s}^\gamma}{u_{n\bk+\bb}^{(0)}}
    &=
    \sum_{p = 1}^{\infty} \mel{u_{m\bk}^{(0)}}{\hat{s}^\gamma}{u_{p\bk}^{(0)}}
    \braket{u_{p\bk}^{(0)}}{u_{n\bk+\bb}^{(0)}} \,,
    \nnnl
    \mel{u_{m\bk}^{(0)}}{\hat{s}^\gamma \hat{H}_\bk}{u_{n\bk+\bb}^{(0)}}
    &=
    \sum_{p = 1}^{\infty} \mel{u_{m\bk}^{(0)}}{\hat{s}^\gamma}{u_{p\bk}^{(0)}}
    \veps_{p\bk} \braket{u_{p\bk}^{(0)}}{u_{n\bk+\bb}^{(0)}} \,.
\end{align}
On the right side of \Eq{eq:Qiao_2} are the usual spin matrix elements and the overlap matrix $M$ [\Eq{eq:M_def}] in the \textit{ab initio} wavefunction basis.
No additional quantities need to be computed.

Although the derivation was exact, there are two potential issues in the practical use of this method.
First, \Eq{eq:Qiao_2} is exact only when all bands are included in the sum over bands $p$.
In practice, one only sums over the $N^{\rm band}(\bk)$ bands that are included in the Wannierization process.
This step is an approximation, although it is accurate in most cases.
Second, \Eq{eq:spin_vel_Qiao_1} makes use of the derivatives of the wavefunction in the eigenstate gauge, which may not be a smooth function of $\bk$.
In case of band crossings or near-degeneracies, terms involving the eigenvector derivative $\partial_\alpha U_{mn\bk} \propto \frac{1}{\veps_\nk - \veps_\mk}$ [Eq.~(33) of Ref.~\cite{2018QiaoSHC}] may be numerically unstable and even diverge.
Note that the spin-velocity itself should always be finite by definition [\Eq{eq:spin_vel_def}].

We now move on to the method of \citet{2019RyooSHC}.
Instead of evaluating the derivative in the eigenstate gauge as in \Eq{eq:spin_vel_Qiao_1}, one can take the derivative in the Wannier gauge and find [see Eq.~(25) of Ref.~\cite{2019RyooSHC}]
\begin{align} \label{eq:sv_Wannier_ryoo}
  &\quad  \hbar \mel{u_{i\bk}^\rmW}{\hat{s}^\alpha \hat{v}_{\beta,\bk}}{u_{j\bk}^\rmW}
  \nnnl
  &= \mel{u_{i\bk}^\rmW}{\hat{s}^\alpha (\partial_\beta  \hat{H}_\bk)}{u_{j\bk}^\rmW}
  \nnnl
  &= \partial_\beta \mel{u_{i\bk}^\rmW}{\hat{s}^\alpha \hat{H}_{\bk}}{u_{j\bk}^\rmW}
  - \mel{\partial_\beta u_{i\bk}^\rmW}{\hat{s}^\alpha \hat{H}_{\bk}}{u_{j\bk}^\rmW}
  \nnnl
  & \quad
  - \mel{u_{i\bk}^\rmW}{\hat{s}^\alpha \hat{H}_{\bk}}{\partial_\beta u_{j\bk}^\rmW}
  \,.
\end{align}
The following quantities are defined in Eqs.~(22) and (23) of Ref.~\cite{2019RyooSHC}:
\begin{align}
  s^{\alpha(\text{W})}_{ij\bk} &= \mel{u_{i\bk}^\rmW}{\hat{s}^\alpha}{u_{j\bk}^\rmW} \,, \\
  \mathcal{S}^{\alpha(\text{W})}_{\beta,ij\bk} &= i\mel{u_{i\bk}^\rmW}{\hat{s}^\alpha}{\partial_\beta u_{j\bk}^\rmW} \,.
\end{align}

Reference~\cite{2019RyooSHC} used the notation $\doteq$ to indicate that the identities hold when the matrix elements are transformed from the Wannier gauge to the Hamiltonian gauge and restricted to the bands inside the frozen (inner) window. Concretely, for the two Wannier matrices $X_{ij\bk}$ and $Y_{ij\bk}$, the equation
\begin{equation} \label{eq:doteq_def1}
    X_{ij\bk} \, \doteq \, Y_{ij\bk}
\end{equation}
means
\begin{multline} \label{eq:doteq_def2}
    \sum_{ij} U_{mi\bk} X_{ij\bk} U^\dagger_{jn\bk}
    = \sum_{ij} U_{mi\bk} Y_{ij\bk} U^\dagger_{jn\bk}
    \\
    \text{ for $\mk, \nk \in \mathcal{W}_{\rm froz}$} \,.
\end{multline}
The dot equality does not imply equality of the full matrix,
since the states outside the frozen window are not accurately represented by the WFs.
Reference~\cite{2019RyooSHC} used the dot equality for the computation of the first and second terms of \Eq{eq:sv_Wannier_ryoo}.
\begin{align} \label{eq:dot1}
    \mel{u_{i\bk}^\rmW}{\hat{s}^\alpha\hat{H}_{\bk}}{u_{j\bk}^\rmW}
    & \doteq \sum_{j'} \mel{u_{i\bk}^\rmW}{\hat{s}^\alpha}{u_{j'\bk}^\rmW} H_{j'j\bk}^\rmW \nnnl
    &=
    [s_{\bk}^{\alpha(\text{W})} H_{\bk}^\rmW]_{ij}
\end{align}
\begin{align} \label{eq:dot2}
    \mel{\partial_\beta u_{i\bk}^\rmW}{\hat{s}^\alpha \hat{H}_{\bk}}{u_{j\bk}^\rmW}
    & \doteq \sum_{j'} \mel{\partial_\beta u_{i\bk}^\rmW}{\hat{s}^\alpha}{u_{j'\bk}^\rmW} H_{j'j\bk}^\rmW \nnnl
    &=
    [i\mathcal{S}_{\beta,\bk}^{\dagger \alpha(\text{W})} H_{\bk}^\rmW]_{ij}
\end{align}
Equations~\eqref{eq:dot1} and~\eqref{eq:dot2} are valid dot equalities since
\begin{align}
    &\quad \sum_{ij} U_{mi\bk} \mel{u_{i\bk}^\rmW}{\hat{s}^\alpha\hat{H}_{\bk}}{u_{j\bk}^\rmW} U^\dagger_{jn\bk} \nnnl
    &=
    \sum_{i} U_{mi\bk} \mel{u_{i\bk}^\rmW}{\hat{s}^\alpha\hat{H}_{\bk}}{u_{n\bk}^\rmH} \nnnl
    &=
    \sum_i U_{mi\bk} \mel{u_{i\bk}^\rmW}{\hat{s}^\alpha}{u_{n\bk}^\rmH} \varepsilon_{n\bk}^\rmH \nnnl
    &=
    \sum_{ij'n'} U_{mi\bk} \mel{u_{i\bk}^\rmW}{\hat{s}^\alpha}{u_{j'\bk}^\rmW} U^\dagger_{j'n'\bk} \varepsilon_{n\bk}^\rmH \delta_{n'n} \nnnl
    &=
    \sum_{ij'j} U_{mi\bk} s_{ij'\bk}^{\alpha\rmW} H_{j'j\bk}^\rmW U_{jn\bk} \nnnl
    &=
    \sum_{ij} U_{mi\bk} [s_{\bk}^{\alpha(\text{W})} H_{\bk}^\rmW]_{ij} U_{jn\bk} \,,
\end{align}
and similarly for \Eq{eq:dot2}.

The problem is that when a derivative is applied to a dot equality $X \doteq Y$, one should take the derivative of \Eq{eq:doteq_def2} as a whole, which gives
\begin{equation}
    \Bigl( \partial_\alpha \bigl[ U (X \!-\! Y) U^\dagger \bigr] \Bigr)_{mn\bk} \!\!= 0 \ \text{if $\mk, \nk \in \mathcal{W}_{\rm froz}$} \,.
\end{equation}
However, this condition does \textit{not} mean that $\partial_\alpha X \doteq \partial_\alpha Y$ holds:
\begin{align}
    &\ \quad
    \bigl[ U (\partial_\alpha X - \partial_\alpha Y) U^\dagger \bigr]_{mm\bk} \nnnl
    &=  \partial_\alpha \bigl[ U (X \!-\! Y) U^\dagger \bigr]
       - U (X \!-\! Y) (\partial_\alpha U^\dagger)
       - (\partial_\alpha U) (X \!-\! Y) U^\dagger
    \nnnl
    &= - U (X \!-\! Y) (\partial_\alpha U^\dagger)
       - (\partial_\alpha U) (X \!-\! Y) U^\dagger
    \nnnl
    &= - U (X \!-\! Y) U^\dagger U (\partial_\alpha U^\dagger)
       - (\partial_\alpha U) U^\dagger U (X \!-\! Y) U^\dagger
    \nnnl
    &\neq 0 \,.
\end{align}
(We omit the $mn\bk$ subscripts after the first line; matrix multiplication is assumed.)
Note that the $U (X \!-\! Y) U^\dagger$ term does not vanish since we assume $X \doteq Y$ but not $X = Y$.
Therefore, even though Eqs.~\eqref{eq:dot1} and~\eqref{eq:dot2} are valid dot equalities, the following equation is \textit{not}:
\begin{equation}
\label{eq:invalid_dot_equality}
    \partial_\beta \mel{u_{i\bk}^\rmW}{\hat{s}^\alpha\hat{H}_{\bk}}{u_{j\bk}^\rmW} \doteq
    \partial_\beta [s_{\bk}^{\alpha(\text{W})} H_{\bk}^\rmW]_{ij} \,.
\end{equation}
It should be corrected as
\begin{align} \label{eq:valid_dot}
    &\quad \partial_\beta \mel{u_{i\bk}^\rmW}{\hat{s}^\alpha\hat{H}_{\bk}}{u_{j\bk}^\rmW}
    \nnnl &\doteq
    \partial_\beta [s_{\bk}^{\alpha(\text{W})} H_{\bk}^\rmW]_{ij}
    \nnnl & + \sum_{j'}\left ( \mel{u_{i\bk}^\rmW}{\hat{s}^\alpha\hat{H}_{\bk}}{u_{j'\bk}^\rmW} - [s_{\bk}^{\alpha(\text{W})} H_{\bk}^\rmW]_{ij'} \right ) [U_\bk^\dagger (\partial_\beta U_\bk)]_{j'j}
    \nnnl & - \sum_{j'} [U_\bk^\dagger (\partial_\beta U_\bk)]_{ij'}\left ( \mel{u_{j'\bk}^\rmW}{\hat{s}^\alpha\hat{H}_{\bk}}{u_{j\bk}^\rmW} - [s_{\bk}^{\alpha(\text{W})} H_{\bk}^\rmW]_{j'j} \right )  \,.
\end{align}
Note that if the dot equality were a real equality, e.\,g.\,, if the Wannier functions come from isolated bands without disentanglement, the additional term in \Eq{eq:valid_dot} vanishes as anticipated.
In practice, the method presented in Ref.~\cite{2019RyooSHC} converges fast with the number of considered electronic bands.

Here, we propose a new method that combines the strengths of the two methods.
Unlike Ref.~\cite{2018QiaoSHC}, this method does not resort to the invalid completeness of the Bloch states in the outer window and is numerically stable, and unlike Ref.~\cite{2019RyooSHC}, this method avoids the use of invalid dot equality.
In this method, we first directly multiply the matrices in the Hamiltonian gauge, not in the Wannier gauge.
\begin{align} \label{eq:revised_dot}
    \mel{u_{i\bk}^\rmW}{\hat{s}^\alpha \hat{H}_{\bk}}{u_{j\bk}^\rmW}
    &= \sum_{mn} V^\dagger_{im\bk} \mel{u_{m\bk}}{\hat{s}^\alpha \hat{H}_{\bk}}{u_n\bk} V_{nj\bk}
    \nnnl
    &= \sum_{mn} V^\dagger_{im\bk} \mel{u_{m\bk}}{\hat{s}^\alpha}{u_{n\bk}} \varepsilon_{n\bk} V_{nj\bk}
\end{align}
We then interpolate from $\mel{u_{i\bk}^\rmW}{\hat{s}^\alpha \hat{H}_{\bk}}{u_{j\bk}^\rmW}$ to $\mel{u_{i\bk}^\rmW}{\hat{s}^\alpha \hat{H}_{\bk}}{u_{j\bk}^\rmW}$.
Also, we can evaluate
\begin{equation} \label{eq:dot2_revised}
    \mel{\partial_\beta u_{i\bk}^\rmW}{\hat{s}^\alpha \hat{H}_{\bk}}{u_{j\bk}^\rmW} = \mel{u_{j\bk}^\rmW}{\hat{H}_{\bk} \hat{s}^\alpha}{\partial_\beta u_{i\bk}^\rmW}^*
\end{equation}
from the Fourier transform of Wannier matrix $\mel{w_{i\bf{0}}}{\hat{H}\hat{s}^\alpha (\hat{r} - R)^\beta}{w_{j\bR}}$, as done in Ref.~\cite{2019RyooSHC}.
Therefore, the Wannier gauge matrix elements needed for the accurate interpolation of the spin-velocity operator are
\begin{gather}
  \mel{u_{i\bk}^\rmW}{\hat{H}_{\bk}}{u_{j\bk}^\rmW} \,,\ %
  \braket{u_{i\bk}^\rmW}{\partial_\beta u_{j\bk}^\rmW} \,, \nnnl
  \mel{u_{i\bk}^\rmW}{\hat{s}^\alpha \hat{H}_\bk}{u_{j\bk}^\rmW} \,, \ %
  \mel{u_{i\bk}^\rmW}{\hat{s}^\alpha \hat{H}_{\bk}}{\partial_\beta u_{j\bk}^\rmW} \,, \nnnl
  \mel{u_{i\bk}^\rmW}{\hat{H}_{\bk} \hat{s}^\alpha}{\partial_\beta u_{j\bk}^\rmW} \,,
  \label{eq:dot2_W_gauge_mtxelts}
\end{gather}
which can be computed from the following real-space Wannier matrix elements:
\begin{gather}
    \mel{w_{i\mb{0}}}{\hat{H}}{w_{j\bR}} \,,\ %
    \mel{w_{i\mb{0}}}{\hat{r}^\beta}{w_{j\bR}} \,, \nnnl
    \mel{w_{i\mb{0}}}{\hat{s}^\alpha}{w_{j\bR}} \,,\ %
    \mel{w_{i\mb{0}}}{\hat{s}^\alpha (\hat{r}-R)^\beta}{w_{j\bR}} \,, \nnnl
    \mel{w_{i\mb{0}}}{\hat{s}^\alpha \hat{H} (\hat{r}-R)^\beta}{w_{j\bR}} \,.
    \label{eq:dot2_real_mtxelts}
\end{gather}

We summarized the three methods for calculating the spin-velocity matrix elements in Table~\ref{tab:spin_vel_methods}.

\begin{figure}[tb]
\centering
\includegraphics[width=0.99\linewidth]{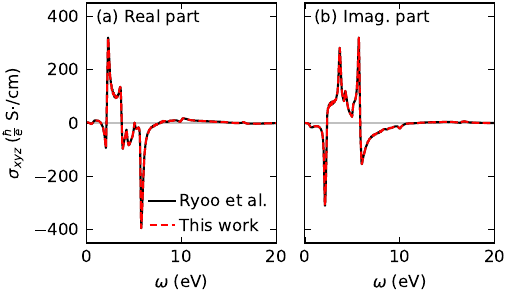}
\caption{
    Real and imaginary parts of the spin Hall conductivity of GaAs with respect to the frequency, where the spin-velocity matrix was calculated using the method of \citet{2019RyooSHC} or the proposed method [Eqs.~\eqref{eq:revised_dot}--\eqref{eq:dot2_real_mtxelts}].
    We used the first-order TEFD method on a $20 \times 20 \times 20$ coarse $\bk$-point grid in both cases.
}
\label{fig:ryoo_sjh_1}
\end{figure}

\begin{figure}[tb]
\centering
\includegraphics[width=0.99\linewidth]{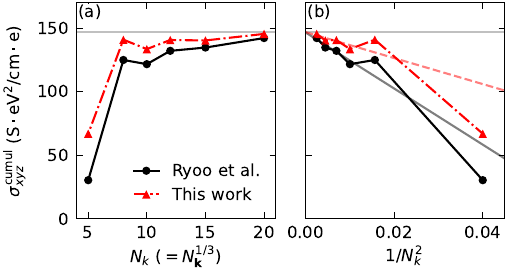}
\caption{
    Cumulative spin Hall conductivity [\Eq{eq:SHC_cumulative}] of GaAs as a function of the (a) coarse grid size and (b) its squared inverse.
    See \Fig{fig:ryoo_sjh_1} for details of the calculation.
}
\label{fig:ryoo_sjh_2}
\end{figure}

Figure~\ref{fig:ryoo_sjh_1} compares the spin Hall conductivity of GaAs computed using the proposed method and the method of Ref.~\cite{2019RyooSHC}.
We find an almost exact agreement between the two methods.
The cumulative spin Hall conductivity, shown in Fig.~\ref{fig:ryoo_sjh_2}, also converges to a similar value as we increase the coarse $\bk$ grid size.
This result means that the correction term in \Eq{eq:valid_dot} is small.
On this basis, we used the method of Ref.~\cite{2019RyooSHC} for the results shown in the main text.

\section{Cumulative spin Hall conductivity} \label{app:cumulative_shc}

In this section, we show that the cumulative spin Hall conductivity defined in \Eq{eq:SHC_cumulative} can be expressed in a more compact form, without the frequency integral.
The Kubo formula for spin Hall conductivity reads~\cite{Yao2005SHC}
\begin{equation}
    \sigma^{\alpha\beta\gamma}(\omega)
    = \frac{e}{\hbar} \nbint{-\infty}{\infty} \dd E \, \frac{N^{\alpha\beta\gamma}(E)}{E^2 - (\hbar\omega + i0^+)^2} \,,
\end{equation}
where
\begin{multline} \label{eq:shc_N_def}
    N^{\alpha\beta\gamma}(E)
    = \sum_{mn\bk} (f_\nk - f_{m\bk}) \, \delta(E - \varepsilon_\nk + \varepsilon_\mk)
    \\
    \times \Im [
        \mel{u_\nk}{j^{\gamma\alpha}_\bk}{u_\mk}
        \mel{u_\mk}{v^\beta_\bk}{u_\nk}
    ]
    \,.
\end{multline}
Then, by using the identity
\begin{align}
    &\Im \frac{1}{E^2 - (\hbar\omega + i0^+)^2}
    \nnnl
    &= \frac{1}{2E} \Im \Bigl(
          \frac{1}{E - \hbar\omega - i0^+}
        + \frac{1}{E + \hbar\omega + i0^+}
    \Bigr)
    \nnnl
    &= \frac{\pi}{2E} \bigl[
        \delta(E - \hbar\omega)
        - \delta(E + \hbar\omega)
    \bigr] \,,
\end{align}
we can write the cumulative spin Hall conductivity [\Eq{eq:SHC_cumulative}] as
\begin{align}
    & \Im  \nbint{0}{\infty} \dd \omega \, \hbar \omega \, \sigma^{\alpha\beta\gamma}(\omega) \,
    \nnnl
    &= \frac{\pi e}{\hbar} \nbint{-\infty}{\infty} \dd E
    \nbint{0}{\infty} \dd \omega \, \hbar \omega \, N^{\alpha\beta\gamma}(E) \frac{\delta(E-\hbar\omega) - \delta(E+\hbar\omega)}{2E}
    \nnnl
    &= \frac{\pi e}{2\hbar^2} \nbint{-\infty}{\infty} \dd E \, N^{\alpha\beta\gamma}(E) \,.
\end{align}
By substituting \Eq{eq:shc_N_def} into this equation, we find
\begin{align}
    & \Im \nbint{0}{\infty} \dd \omega \, \hbar \omega \, \sigma^{\alpha\beta\gamma}(\omega) \,
    \nnnl
    &= \frac{\pi e}{2\hbar^2}\sum_{mn\bk} (f_\nk - f_\mk)
    \Im [\mel{u_\nk}{j^{\gamma\alpha}_\bk}{u_\mk}
    \mel{u_\mk}{v_\bk^\beta}{u_\nk}]
    \nnnl
    &= \frac{\pi e}{2\hbar^2} \Im \sum_{mn\bk} \Bigl[
        f_\nk \mel{u_\nk}{j^{\gamma\alpha}_\bk}{u_\mk} \mel{u_\mk}{v_\bk^\beta}{u_\nk}
    \nnnl
    &\hspace{6em} - f_\mk \mel{u_\mk}{v_\bk^\beta}{u_\nk} \mel{u_\nk}{j^{\gamma\alpha}_\bk}{u_\mk}
    \Bigr]
    \nnnl
    &= \frac{\pi e}{2\hbar^2} \Im \sum_{n\bk} f_\nk \mel{u_\nk}{[ j^{\gamma\alpha}_\bk, v_\bk^\beta ]}{u_\nk} \,.
\end{align}
This final expression is fully determined by the eigenstates and eigenenergies of the occupied states.
Thus, they can be computed directly without integrating over frequencies, although we did not do so in this work.

\section{Detailed derivations for higher-order finite difference}
\label{app:derivation_hofd}

\subsection{HOFD conditions and the error}
We derive the HOFD condition in \Eq{eq:fd_nearest_nd}.
First, we begin by reviewing the first-order case.
For a set of $\bb$ vectors that connect $\bk$ to neighboring grid points, we approximate the $\bk$-gradient with the ansatz
\begin{equation} \label{eq:nd_ansatz}
    \nabla_\alpha f(\bk)
    \approx \nabla_\alpha^{\rm FD} f(\bk)
    = \sum_{\bb}  c_{\abs{\bb}} b^\alpha f(\bk + \bb) \,.
\end{equation}
Now, let us Taylor expand $f(\bk +\bb)$ around $\bk$ as
\begin{equation}
\label{eq:Taylor_expansion_k_b}
    f(\bk + \bb) = \sum_{n=0}  \frac{f_{\ai \aii \cdots \an}^{(n)}}{n!} b^{\ai}b^{\aii}\cdots b^{\an}\,,
\end{equation}
where summation over repeated Cartesian indices $\alpha_1, \cdots, \alpha_n$ is implied.
Here, $f_{\ai \aii \cdots \an}^{(n)}$ denotes the $n$th-order partial derivative of $f(\bk)$.
If we combine the above two equations, we get
\begin{align}
\label{eq:nd_ansatz_first_taylor}
    &\nabla_\alpha^{\rm FD} f(\bk)
    \\
    &= \sum_{\bb} c_{\abs{\bb}} b^\alpha \sum_{n=0}  \frac{f_{\ai \aii \cdots \an}^{(n)}}{n!} b^{\ai}b^{\aii}\cdots b^{\an}
    \nnnl
    &= \sum_{n=0}\frac{f_{\ai \aii \cdots \an}^{(n)}}{n!} \sum_{\bb} c_{\abs{\bb}} b^\alpha b^{\ai}b^{\aii}\cdots b^{\an}
    \nnnl
    &= \sum_{n=1,3,5,...}\frac{f_{\ai \aii \cdots \an}^{(n)}}{n!} \sum_{\bb} c_{\abs{\bb}} b^\alpha b^{\ai}b^{\aii}\cdots b^{\an} \nnnl
    &\hspace{15em}
    \ (\,\because\, c_{\abs{-\bb}}=c_{\abs{\bb}}\,)
    \nnnl
    &=f_{\ai}^{(1)}\sum_\bb c_{\abs{\bb}} b^\alpha b^{\ai}  + f_{\ai\aii\alpha_{3}}^{(3)}\sum_\bb c_{\abs{\bb}}b^\alpha b^{\ai}b^{\aii}b^{\alpha_{3}} + \cdots \,.
    \nonumber
\end{align}
In the last line of \Eq{eq:nd_ansatz_first_taylor}, the first term corresponds to the desired first-order derivative, and the remainder is the FD error.
For this equation to correctly approximate $f^{(1)}_\alpha$, we need the condition \Eq{eq:c_b_sum} to be satisfied, which we repeat here for convenience:
\begin{equation}
    \sum_\bb c_{\abs{\bb}} b^\alpha b^{\ai} = \delta_{\alpha \ai} \,.
\end{equation}
From this condition, it follows that the weights scale as $c_{\abs{\bb}}\sim O(b^{-2})$.
Therefore, the leading error of the first-order FD approximation is quadratic:
\begin{equation}
    \abs{\nabla_\alpha^{\rm FD} f(\bk) - f_{\alpha}^{(1)}} \sim c_{\abs{\bb}} O(b^4) \sim O(b^2) \,.
\end{equation}

For the HOFD method, we impose additional constraints so that more FD error terms in the last line of \Eq{eq:nd_ansatz_first_taylor} vanish.
For the leading FD error proportional to $f^{(3)}_{\alpha_1 \alpha_2 \alpha_3}$ to vanish, we need
\begin{equation} \label{eq:cond_2}
    \sum_\bb c_{\abs{\bb}} b^\alpha b^{\ai} b^{\aii} b^{\alpha_3}= 0 \,.
\end{equation}
The second-order FD method imposes this constraint.
Then, the leading FD error comes from $f^{(5)}$ and scales as $O(b^4)$.
Generalizing this idea to $N$th-order HOFD, which requires $N-1$ FD error terms to vanish, we obtain the condition in \Eq{eq:fd_nearest_nd}.
If all these conditions are met, the $N$th-order HOFD error comes from $f^{(2N+1)}$ and scales as $O(b^{2N})$:
\begin{equation}
\label{eq:hofd_error}
    \abs{\nabla_\alpha^{\rm FD} f(\bk) - f_{\alpha}^{(1)}} \sim c_{\abs{\bb}} O(b^{2(N+1)}) \sim O(b^{2N}) \,.
\end{equation}

\subsection{Multiples method}

In this subsection, we derive the $c_{\abs{\bb}}$ coefficients for the HOFD multiples method.
We first review the one-dimensional case and then discuss how this can be combined with the first-order FD method to obtain HOFD methods in higher dimensions.

In the one-dimensional $N$th-order HOFD method, $\bb$ vectors are taken from the set $\{-Nb, -(N-1)b, \cdots, -b, b, \cdots, Nb\}$, and the condition for $c_{\abs{\bb}}$'s in \Eq{eq:fd_nearest_nd} becomes
\begin{align} \label{eq:eqs_1d}
    \sum_{m=-N}^{N} c_{mb}^{(N)} (m b)^2 &= 1 \,,
    \nnnl
    \sum_{m=-N}^{N} c_{mb}^{(N)} (m b)^4 &= 0 \,,
    \nnnl
    &\ \vdots
    \nnnl
    \sum_{m=-N}^{N} c_{mb}^{(N)} (m b)^{2N} &= 0 \,.
\end{align}
Using the even parity of the weights, $c^{(N)}_{-mb} = c^{(N)}_{mb}$, we can simplify these equations as
\begin{align} \label{eq:eqs_1d_2}
\begin{split}
    c_b^{(N)} + 2^2 c_{2b}^{(N)} + \cdots + N^2 c_{Nb}^{(N)} &= \frac{1}{2b^2} = c_b^{(1)} \\
    c_b^{(N)} + 2^4 c_{2b}^{(N)} + \cdots + N^4 c_{Nb}^{(N)} &= 0 \\
    &\qquad \vdots \\
    c_b^{(N)} + 2^{2N} c_{2b}^{(N)} + \cdots + N^{2N} c_{Nb}^{(N)} &= 0\,,
\end{split}
\end{align}
and write them in a matrix form:
\begin{equation} \label{eq:vandermonde}
    \begin{pmatrix}
      1^{2} & 2^{2} & \cdots & N^{2}\\
      1^{4} & 2^{4} & \cdots & N^{4}\\
      \vdots&\vdots & \ddots & \vdots\\
      1^{2N} & 2^{2N} & \cdots & N^{2N}\\
    \end{pmatrix}
    \begin{pmatrix}
      c_{b}^{(N)} \\
      c_{2b}^{(N)} \\
      \vdots \\
      c_{Nb}^{(N)}
    \end{pmatrix}
    =
    \begin{pmatrix}
      c_{b}^{(1)} \\
      0 \\
      \vdots \\
      0
    \end{pmatrix}\,.
\end{equation}
The $N\times N$ matrix on the left is the Vandermonde matrix, whose inverse has a known analytic form.
Solving \Eq{eq:vandermonde}, we obtain the $N$th-order HOFD coefficients:
\begin{equation} \label{eq:cramers}
    c_{mb}^{(N)} = c_b^{(1)} \frac{1}{m^2} \prod_{n \neq m}^{N} \frac{n^2}{n^2 - m^2}\,.
\end{equation}
%

Now, we show that these one-dimensional HOFD coefficients can be used to compute higher-dimensional HOFD coefficients.
Given the first-order FD vectors $\{ \bb \}$ and coefficients $c^{(1)}_\bb$'s, we choose the $N$th-order FD vectors as $\{ \bb \} \cup \{ 2\bb\} \cup \cdots \cup \{ N \bb \}$, and the coefficients as
\begin{equation} \label{eq:cramers_nd}
    c_{\abs{m\bb}}^{(N)} = c_{\abs{\bb}}^{(1)} \frac{1}{m^2} \prod_{n \neq m}^{N} \frac{n^2}{n^2 - m^2} \,,
\end{equation}
using the same factor as in the one-dimensional case [\Eq{eq:cramers}].
To show that this choice satisfies the HOFD condition, we put \Eq{eq:cramers_nd} into the left side of \Eq{eq:fd_nearest_nd}:
\begin{align}
\begin{split}
\label{eq:fd_nearest_nd_proof}
    \sum_{\bb} b^\ai b^\aii & \left[c_{\abs{\bb}}^{(N)} + 2^2 c_{\abs{2\bb}}^{(N)} + \cdots + N^2 c_{\abs{N\bb}}^{(N)}\right] \\
    &= \sum_{\bb} b^\ai b^\aii c_{\abs{\bb}}^{(1)} = \delta_{\ai \aii}\\
    \sum_{\bb} b^\ai b^\aii & b^{\alpha_3} b^{\alpha_4} \left[c_{\abs{\bb}}^{(N)} + 2^4 c_{\abs{2\bb}}^{(N)} + \cdots + N^4 c_{\abs{N\bb}}^{(N)}\right] = 0 \\
    &\qquad \vdots \\
    \sum_{\bb} b^\ai b^\aii & \cdots b^{\alpha_{2N}} \\
    &\times \left[c_{\abs{\bb}}^{(N)} + 2^{2N} c_{\abs{2\bb}}^{(N)} + \cdots + N^{2N} c_{\abs{N\bb}}^{(N)}\right] = 0\,,
\end{split}
\end{align}
since $c_{\abs{m\bb}}^{(N)}$'s are, by construction, also the solution of \Eq{eq:eqs_1d_2}.
The last equality in the first line is from the first-order FD condition, \Eq{eq:c_b_sum}.
We thus proved that $c_{\abs{m\bb}}^{(N)}$'s [Eq.~\eqref{eq:cramers_nd}] satisfy the HOFD condition [\Eq{eq:fd_nearest_nd}].

\FloatBarrier 

\bibliography{main}

\end{document}